\documentclass[aps,prb,twocolumn,superscriptaddress, floatfix]{revtex4}

\usepackage{graphicx}
\usepackage{url}
\usepackage{natbib}

\newcommand{\TBCO}{Tl$_2$Ba$_2$CuO$_{6 \pm \delta}$}
\newcommand{\BSCCO}{${\rm Bi_2Sr_2CaCu_2O_8}$}
\newcommand{\YBCO}{YBa$_2$Cu$_3$O$_{6+y}$}
\newcommand{\Overdoped}{YBa$_2$Cu$_3$O$_{6.993}$}
\newcommand{\OrthoII}{YBa$_2$Cu$_3$O$_{6.50}$}
\newcommand{\GYBCO}{Gd$_{\rm x}$Y$_{\rm 1-x}$Ba$_2$Cu$_3$O$_{6+y}$}

\newcommand{\dwave}{{\it d}-wave}
\newcommand{\cuplane}{CuO$_2$}

\begin{document}
\title{Phenomenology of $\hat a$-axis and $\hat b$-axis charge dynamics from microwave spectroscopy of highly ordered \OrthoII\ and \Overdoped}

\author{R.~Harris}
\affiliation{Department of Physics and Astronomy,
University of British Columbia, 6224 Agricultural Road, Vancouver, British Columbia, Canada V6T 1Z1}
\affiliation{D-Wave Systems Inc., 4401 Still Creek Drive, Burnaby, British Columbia, Canada, V5C 6G9}
\author{P.~J.~Turner}
\affiliation{Department of Physics and Astronomy,
University of British Columbia, 6224 Agricultural Road, Vancouver, British Columbia, Canada V6T 1Z1}
\affiliation{Department of Physics, Simon Fraser University, 8888 University Drive, Burnaby, British Columbia, Canada, V5A 1S6}
\author{Saeid~Kamal}
\affiliation{Department of Physics and Astronomy,
University of British Columbia, 6224 Agricultural Road, Vancouver, British Columbia, Canada V6T 1Z1}
\author{A.~R.~Hosseini}
\affiliation{Department of Physics and Astronomy,
University of British Columbia, 6224 Agricultural Road, Vancouver, British Columbia, Canada V6T 1Z1}
\author{P.~Dosanjh}
\affiliation{Department of Physics and Astronomy,
University of British Columbia, 6224 Agricultural Road, Vancouver, British Columbia, Canada V6T 1Z1}
\author{G.~K.~Mullins}
\affiliation{Department of Physics and Astronomy,
University of British Columbia, 6224 Agricultural Road, Vancouver, British Columbia, Canada V6T 1Z1}
\affiliation{School of Engineering Science, Simon Fraser University, 8888 University Drive, Burnaby, British Columbia, Canada, V5A 1S6}
\author{J.~S.~Bobowski}
\affiliation{Department of Physics and Astronomy,
University of British Columbia, 6224 Agricultural Road, Vancouver, British Columbia, Canada V6T 1Z1}
\author{C.~P.~Bidinosti}
\affiliation{Department of Physics and Astronomy,
University of British Columbia, 6224 Agricultural Road, Vancouver, British Columbia, Canada V6T 1Z1}
\affiliation{Department of Physics, Simon Fraser University, 8888 University Drive, Burnaby, British Columbia, Canada, V5A 1S6}
\author{D.~M.~Broun}
\affiliation{Department of Physics and Astronomy,
University of British Columbia, 6224 Agricultural Road, Vancouver, British Columbia, Canada V6T 1Z1}
\affiliation{Department of Physics, Simon Fraser University, 8888 University Drive, Burnaby, British Columbia, Canada, V5A 1S6}
\author{Ruixing~Liang}
\author{W.~N.~Hardy}
\author{D.~A.~Bonn}
\affiliation{Department of Physics and Astronomy,
University of British Columbia, 6224 Agricultural Road, Vancouver, British Columbia, Canada V6T 1Z1}
\date{\today}

\begin{abstract}

Extensive measurements of the microwave conductivity of highly pure
and oxygen-ordered \YBCO\ single crystals have been performed as a
means of exploring the intrinsic charge dynamics of a $d$-wave
superconductor. Broadband and fixed-frequency microwave apparatus
together provide a very clear picture of the electrodynamics of the
superconducting condensate and its thermally excited nodal
quasiparticles. The measurements reveal the existence of very
long-lived excitations deep in the superconducting state, as
evidenced by sharp cusp-like conductivity spectra with widths that
fall well within our experimental bandwidth. We present a
phenomenological model of the microwave conductivity that captures
the physics of energy-dependent quasiparticle dynamics in a $d$-wave
superconductor which, in turn, allows us to examine the scattering
rate and oscillator strength of the thermally excited quasiparticles
as functions of temperature. Our results are in close agreement with
the Ferrell-Glover-Tinkham sum rule, giving confidence in both our
experiments and the phenomenological model.  Separate experiments
for currents along the $\hat a$ and $\hat b$ directions of detwinned crystals allow
us to isolate the role of the CuO chain layers in \YBCO, and a model
is presented that incorporates both  one-dimensional conduction from
the chain electrons and two-dimensional transport associated with
the \cuplane\ plane layers.

\end{abstract}

\maketitle
\section{Introduction}

Microwave conductivity measurements have provided a powerful
method for investigating the low energy charge dynamics of high
temperature cuprate superconductors.  Early on, the observation of a
linear temperature dependence of the magnetic penetration depth
$\lambda(T)$ in YBa$_2$Cu$_3$O$_{6.95}$ provided some of the first
evidence of $d$-wave pairing in these materials.\cite{Hardy1} Over a
decade later, the $d$-wave superconducting phase remains one of the
cornerstones of our understanding of the cuprate problem.
Surprisingly, the physics of the superconducting state seems to be
well described as a BCS instability of a two-dimensional Fermi
liquid.  However, it is essential that this conjecture be thoroughly
tested, and one of the best ways to do this is through detailed
measurements of charge transport in high quality single crystals. In
this work we report on microwave spectroscopy of two gold-standard
cuprate materials --- Ortho-I \Overdoped\ and Ortho-II \OrthoII\ ---
and develop a simple phenomenology that describes their low-energy
charge dynamics.

Understanding the role of impurities in the cuprate superconductors
has taken many years of great effort.   The materials have
complicated chemistry --- containing at least four elements --- and have
much  greater sensitivity to disorder than conventional
superconductors due to finite-angular-momentum Cooper pairing.
Early measurements on the temperature dependence of the magnetic penetration depth of \YBCO\ showed deviations from the linear temperature dependence expected for a clean $d$-wave superconductor.  It was later suggested that these effects could be the result of pair-breaking disorder,\cite{Annett,Scalapino} a conjecture that was
confirmed by showing that the controlled addition of Zn impurities
to high quality \YBCO\ crystals caused a crossover in $\Delta\lambda(T)$
from $T$ to $T^2$ behaviour.\cite{BonnZn,achkir93, BonnCzech}   Ongoing efforts
to uncover the intrinsic physics of the cuprates depend on continued
improvements in sample quality. The samples used in this work are
the most well-ordered cuprates available, and are of such exceedingly high
purity and crystallinity that even the off-plane dopant oxygen
atoms are highly-ordered.\cite{Ruixing, Ruixing2,Ruixing3}

Some of the earliest measurements of the low energy electromagnetic
response of the cuprates were made by Nuss {\em et al.}, who
measured the conductivity of YBa$_2$Cu$_3$O$_{7}$~ thin films using
terahertz spectroscopy over the frequency range 0.5~THz to
2.5~THz.\cite{Nuss}  The observation of a peak in $\sigma_1(T)$
below the critical temperature $T_c$ was attributed to a strong
decrease in inelastic scattering, rather than the conductivity
coherence peak that occurs in $s$-wave BCS superconductors. Shortly
afterward, Bonn {\em et al.} measured a similar
peak in $\sigma_1(T)$ in the microwave range.\cite{Bonn92,Bonn93} They
proposed that this was the result of a competition between the rapid
increase in quasiparticle lifetime $\tau$ and the decaying
quasiparticle density $n_n(T)$ as the temperature was reduced. This
analysis was carried out within the context of a Drude model
characterized by an energy-independent transport relaxation rate $\tau^{-1}$.
Other transport measurements including thermal conductivity
\cite{yu92} and higher frequency electromagnetic measurements
\cite{romero92} have also confirmed the rapid collapse in scattering
on entering the superconducting state.

In a more recent publication, our group presented a series of
microwave measurements that confirmed and extended these earlier
findings.\cite{Hosseini1} This was made possible by two experimental
developments: the advent of higher purity \YBCO\ crystals grown in
BaZrO$_3$ crucibles and the development of a sufficient number of
fixed-frequency cavity perturbation systems to map out a coarse
microwave conductivity spectrum. Five superconducting resonators,
operating between 1~GHz and 75~GHz, were used to examine the surface
impedance of {\em the same single crystal sample} of fully-doped
YBa$_2$Cu$_3$O$_{6.993}$ over the temperature range 2~K to 100~K (the
conductivity inferred from these data is shown in
Fig.~\ref{fig:figure6} of the present article). At each temperature
the conductivity spectrum was fit to a simple Drude model, which
captured the features of the data well, thus supporting the assumptions
used in the earlier work. The temperature dependent quasiparticle
scattering rate $\tau^{-1}$ derived from these fits was seen to
decrease rapidly below $T_c$ and saturate at a constant
value of 6$\times 10^{10}$~s$^{-1}$ below 20~K,
corresponding to a spectral width of approximately 9~GHz. Assuming a
Fermi velocity $v_F=2\times10^7$cm/s,\cite{Zhou} the residual
scattering rate implies a quasiparticle mean free path of 4~$\mu$m,
a very large distance for a complicated quaternary oxide.   

However, the observation of a temperature independent transport relaxation rate is at
odds with standard models for the electrical conductivity of a
$d$-wave superconductor\cite{Berlinsky}  and resolving this elastic
scattering problem is a major goal of the work presented in this article. The
strong temperature dependence of the transport relaxation rate at higher
temperatures, where the scattering processes are inelastic, has
motivated models in which the opening of a gap in the spectrum of
electronic excitations in the system leads to a collapse of the
inelastic transport relaxation rate.\cite{walker00,duffy01} Finally, the same
set of fixed frequency resonators was used to probe the conductivity
for currents along the crystal $\hat b$-axis of Ortho-I
\Overdoped.  In these measurements, currents are parallel to the CuO
chains and reveal an anisotropy in charge conduction that can be
attributed to the sum of a  narrow \cuplane\ plane quasiparticle
spectrum and a very broad spectral feature arising from the
one-dimensional CuO chains.\cite{Harris1}

Greater insight into the low temperature charge dynamics
of \YBCO\ was gained following the development of a novel low
temperature microwave spectrometer that could measure the
conductivity spectrum as a {\em continuous} function of
frequency.\cite{TurnerPRL, TurnerRSI} Spectroscopy of Ortho-II
\OrthoII\ revealed that the nodal quasiparticle spectrum {\em did}
in fact contain most of the features expected for weak-limit
impurity scattering in a clean $d$-wave superconductor, namely a
sharp cusp-like spectrum and a linear-in-temperature spectral width.
In contrast, similar spectroscopy of Ortho-I \Overdoped\ material
displayed much more Drude-like spectra, except at the very lowest
temperatures.  We return to examining these features in the present
article, presenting a complete body of microwave data on \mbox{Ortho-I} and
Ortho-II \YBCO, and show how these data can be described by a simple
phenomenlogy that captures the physics of energy-dependent charge
dynamics.

The paper is organized as follows.  Section~\ref{sec:experiment} begins
with a brief discussion of the two main experimental techniques,
fixed-frequency microwave cavity perturbation and broadband
bolometric surface resistance spectroscopy.  In
Sec.~\ref{sec:phenomenology} we use the broadband bolometric surface
resistance data to extract detailed conductivity spectra and present
a simple phenomenological model of the conductivity.  With this in
hand, in Sec.~\ref{sec:conductivity} we process data from the five
fixed-frequency experiments, which provide a means of capturing the
conductivity spectrum at higher temperatures than can be reached
with the bolometric experiment.  Finally, we examine the two key
parameters that emerge from fitting to the experimental data:  the
width of the conductivity spectrum, which measures the thermally-averaged transport relaxation rate (Sec.~\ref{subsec:spectralwidth}) and the total integrated spectral
weight or `quasiparticle oscillator strength'
(Sec.~\ref{subsec:spectralweight}).  Throughout this article we use
recently revised absolute values of the $T\rightarrow0$ penetration depths
obtained using a newly developed zero-field ESR absorption
method.\cite{TamiPRB}  The revised penetration depths are
considerably shorter than existing values in the literature and
result in substantial quantitative changes to the previously
published quasiparticle conductivity and superfluid density.
However, we emphasize that neither the qualitative form of the
conductivity data nor the main conclusions drawn from it are
affected.

\section{Experimental Techniques}
\label{sec:experiment}

In this paper we focus on the microwave conductivity, $\sigma=\sigma_1-\textrm{i} \sigma_2$, in the long wavelength limit and in the linear response regime.  In experiments that probe the conductivity, the quantity that is inferred from measurements is the microwave surface impedance $Z_s=R_s+{\rm i}X_s$.  When the applied microwave fields are screened on a length scale that is large compared to the characteristic size of the wavefunctions of both the Cooper pairs and the single-particle excitations (the coherence length $\xi$ and mean free path $\ell$, respectively) it is possible to describe the charge response by local electrodynamics, in which case:
\begin{equation} \label{eqn:Zs}
Z_s=R_s+\textrm{i}X_s=\sqrt{ {\textrm{i} \omega \mu_{\rm 0}} \over
{\sigma} }.
\end{equation}
In very clean cuprate samples the mean free path will exceed the
penetration depth  and the electrodynamics will not necessarily
remain in the local limit.  However, it is possible to take
advantage of the two dimensional electronic structure of the
cuprates, which gives rise to a short coherence length and mean free
path along the crystal $\hat c$ direction, to ensure that the charge
response {\em is} local.\cite{Kosztin-Leggett, Chang-Scalapino}  In
the work reported here we have exclusively used such
geometries.

As pairs condense in a superconductor, the reactive response of the superfluid
quickly overwhelms the dissipative response of the quasiparticle
excitations, and over most of the temperature range
$\sigma_2\gg\sigma_1$.  The local electrodynamic relation then
simplifies to the following approximate forms:
\begin{eqnarray}\label{eqn:RX}
R_s(\omega,T)&\simeq&{1 \over 2} \mu_{\rm 0}^2\omega^2\lambda^3(T)\sigma_1(\omega,T),\\
\nonumber
X_s(\omega,T)&\simeq&\mu_{\rm 0}\omega\lambda(T).
\end{eqnarray}
From these expressions it is clear that a measurement of the surface resistance
$R_s$ contains information about the real part of the conductivity
$\sigma_1$, while a measurement of the surface reactance $X_s$ is
a direct probe of the penetration depth $\lambda$.  As with nearly
all forms of spectroscopy, technical limitations require one to
work with incomplete spectroscopic information and subsequently to rely on causality
to extract the quantities of interest.  Our present work is no
exception. However, in this article we present a wide range
of measurements that not only allow us to extract the
frequency-dependent conductivity, but also provide a set of checks and
balances through the oscillator-strength sum rule that let us verify
our results in detail.
 
In these measurements, a well-defined electromagnetic geometry must
be used for three reasons: to separate the individual components of
the conductivity tensor, to ensure that demagnetization effects are
well controlled, and to enforce local electrodynamics. A
particularly clean approach that has been widely used is to place the
sample in a microwave enclosure near a position of high symmetry, where there is an electric node and the microwave magnetic field is quasi-homogeneous.

Single crystal samples of many of the cuprate superconductors grow
naturally as platelets having a broad $\hat{a}$--$\hat{b}$ plane
crystal face and thin $\hat{c}$-axis dimension. In this case,
demagnetization effects are minimized if the broad face of the
sample is aligned parallel to the microwave magnetic  field. In
response to the microwave fields, screening currents flow near the
surface of the sample, across the broad $\hat{a}$--$\hat{b}$ face,
closing along the $\hat{c}$ direction to complete a loop. All
measurements presented in this article employ thin crystals with
negligible $\hat c$-axis contribution, and were carried out in this
low-demagnetizing factor orientation.

An experimental feature common to each of the seven sets of
apparatus employed in this work is the use of a sapphire hot
finger.\cite{HotFinger}  This arrangement allows a sample to be held
at an elevated temperature in the microwave fields, while the
resonator remains at base temperature. For the field configuration we utilize, the sapphire plate is transparent to microwaves and makes a negligible contribution to the
dissipation. The sample is affixed to the sapphire plate (with a very small amount of silicone grease) that is then inserted into the microwave field through a hole in the
metal end-wall that is small enough to cut off microwave propagation
at the operating frequency.  The thermometer and heater are located
outside the cut-off hole, in intimate thermal contact with the
sapphire isothermal stage, which is in turn weakly linked to the
1.2~K pumped helium bath.

Resolving the microwave surface impedance of small single-crystal cuprate samples over a wide temperature range has been achieved by the use of cavity-perturbation techniques.\cite{Bonn-Hardy, Ormeno} In these experiments the sample
under test is brought into the microwave fields of a high
quality-factor (high-$Q$) resonant structure such as a
superconducting cavity or a loop--gap resonator.  In cavity perturbation, shifts in the resonant frequency $f$ and quality factor $Q$ are related to the surface impedance via:\cite{Waldron}
\begin{equation}
\label{eqn:cavitypert} 
{{\delta f} \over {f}} - {\rm i} {{1} \over {2}} \delta \! \left( {{1}\over{Q}} \right) = {\rm i} \Gamma \Delta Z_s = - \Gamma (\Delta X_s- {\rm i} R_s).
\end{equation}

The quantity $\delta (1/Q)=1/Q^{\rm load}(T)-1/Q^{\rm unload}$, where $Q^{\rm load}(T)$ is the quality factor of the resonator when it contains the sample at temperature $T$, and $Q^{\rm unload}$ is the quality factor of the empty resonator.  $\delta f=f^{\rm load}(T)-f^{\rm load}(T\!=\!1.2~{\rm K})$ is the shift in frequency of the loaded resonator on warming the sample from base temperature to temperature $T$.  Similarly, $\Delta X_s=X_s(T)-X_s(T\!=\!1.2~{\rm K})$.  $R_s$ is the absolute surface resistance, and the geometric factor $\Gamma$ depends on the dimensions of both the cavity and the sample, as well as the geometry of the fields that are being perturbed, and is determined empirically.  

The direct relationship in Eq.~\ref{eqn:cavitypert} assumes that the sample is much thicker than the length scale of the electromagnetic screening --- either the superconducting penetration depth $\lambda$ or the normal-state skin depth. In situations when the screening currents penetrate to a depth approaching the sample thickness, Eq.~\ref{eqn:cavitypert} must be modified.  For our \YBCO\ samples, which have a thickness $\approx 10~\mu m$, this becomes important only in a very narrow temperature region just below $T_c$, and in the normal state where the skin depth is large.  To account for this effect, we use a model that solves the electromagnetic field equations in our one-dimensional geometry with the microwave field applied parallel to the broad face of the sample.\cite{KamalPhD} The result is a more complete version of Eq.~\ref{eqn:cavitypert} that explicitly involves the sample thickness $t$:
\begin{equation}
\label{eqn:cavitypert2} 
{{\delta f} \over {f}} - {\rm i} { {1}\over{2}}  \delta \! \left( {{1}\over{Q}} \right) =\beta {{V_s}\over{V_c}} \left[ 1 - {{\tanh(\kappa t/2)}\over{\kappa t/2}} \right]
\end{equation}
where $V_s$ is the sample volume, $V_c$ is the effective cavity volume, $\beta$ is a constant set by the type of cavity and its field configuration, and $\kappa=\sqrt{{\rm i} \omega \mu_{\rm 0} \sigma}$ is the complex propagation constant for the electromagnetic field in the sample.  For this expression, $\delta f$ is modified and is now the shift in frequency on inserting the sample into the empty resonator, $\delta f=f^{\rm load}(T)-f^{\rm unload}$.  The physical quantities of interest, including $R_s$, $X_s$, $\lambda$, $\sigma_1$ and $\sigma_2$ are determined from measurements of $\delta f$ and $\delta (1/Q)$ and the relationships $\sigma=\kappa^2/({\rm i} \mu_{\rm 0} \omega)$ and $Z_s={\rm i} \mu_{\rm 0} \omega/\kappa$.  

While it is possible to calibrate $R_s$ measurements to very good accuracy, cavity-perturbation techniques do not easily yield an absolute value for
$\lambda(T \to 0)$ since a measurement of the absolute penetration depth amounts to a comparison of the total volume of the sample to the volume of the sample minus the small field-penetrated region at the surface of the sample.  It is
impossible to measure the dimensions of a single crystal
sample precisely enough to achieve an absolute measure of the
in-plane $\lambda$ this way.  Recent Gd ESR measurements have
provided a breakthrough in absolute penetration depth measurements
by using the characteristic microwave absorption of Gd ions lightly
doped through the crystal as a magnetic tracer to measure the field-penetrated volume
at the surface of the sample {\em directly}.\cite{TamiPRB}

The calibration procedure for the superconducting resonators relies on measurements of well-characterized PbSn samples of similar dimensions as the \YBCO\ crystal, as well as runs without any sample at all. These measurements allow for the small corrections necessary to account for background losses and the finite volume of the crystal. Each resonator experiment must be calibrated independently. As a result, there is up to a 10\% systematic uncertainty in the overall scale factor at each frequency.  Stochastic noise is typically much less than 5\% for measurements well below $T_c$.
  
In cavity perturbation experiments, the resonator is generally
restricted to operate at a single fixed frequency, requiring the use
of many separate sets of experimental apparatus to capture a
frequency spectrum. Furthermore, the cavity perturbation method
requires that the dissipation of the unknown sample be comparable to
the dissipation of the cavity itself in order for the absorption to
be measured with high precision.  This is a strong requirement on
the cavity when the sample is a high quality superconductor in the
$T\to0$ limit. The measurement of the residual absorption in
superconductors is challenging at any frequency: in the case of
infra-red spectroscopy the problem becomes that of measuring values
of reflectance that are very close to unity. In microwave cavity
perturbation measurements there is a similar problem of discerning
the small difference in cavity Q with the sample in and out. Both
techniques have problems with calibration and discerning small
differences between two large numbers.

These sensitivity problems are circumvented in a direct absorption
measurement. We have recently developed an apparatus capable of
obtaining continuous-frequency measurements of the absolute
microwave surface resistance in low-loss single crystal samples,
which has been described in detail elsewhere.\cite{TurnerRSI} The
instrument employs a bolometric method of detection in which the
sample of interest is exposed to a microwave frequency magnetic
field $H_{\rm rf}$ whose frequency can be varied, and the
corresponding temperature rise is measured. This method relies on
the fact that the power absorption in a microwave magnetic field is
directly proportional to the surface resistance $R_s$:
\begin{equation}
\label{eqn:power} \ P_{\rm abs}= R_s \int_S H_{\rm rf}^2 \textrm{d}S,
\end{equation}
where $H_{\rm rf}$ is the root-mean-square magnitude of the magnetic
field at the surface $S$. Measurements of the temperature rise of
the sample as a function of frequency directly give $R_s(\omega)$.
To enhance rejection of spurious temperature variations, the rf
power is amplitude modulated at low frequency and the resulting
temperature oscillations of the sample are detected synchronously.
To provide a frequency-tunable microwave magnetic field, the sample
is placed in a custom-made rectangular coaxial transmission line
that supports a TEM mode in which the $H$ fields lie in the
transverse plane and form closed loops around the centre conductor.
The line is terminated by shorting the center conductor and outer
conductor with a flat, metallic endwall, thus enforcing an electric
field node at the end of the waveguide.  The sample is positioned
with its flat face parallel and very close to the endwall such that
it experiences spatially uniform fields over its dimensions.  A
second metallic reference sample of known absorption is placed in an
electrically equivalent position that serves to calibrate the
absolute field strength, which varies considerably across the
frequency range due to standing waves in the transmission line. This
technique has provided a very sensitive means of measuring the
broadband surface resistance spectrum in superconducting crystals
over the range 0.5~GHz to 21~GHz, permitting the first exploration
of the detailed low temperature conductivity line shape in a number
of compounds.\cite{TurnerPRL, SibelCeCoIn5, DavePrOs4Sb12}  The
conductivity spectrum $\sigma(\omega)$ can be extracted very simply
from $R_s(\omega)$ using Eq.~\ref{eqn:RX}. Our full analysis is
slightly more detailed, allowing for the small frequency dependence
of the penetration depth that comes from thermally excited
quasiparticles.

\begin{figure}
\includegraphics[width=\columnwidth]{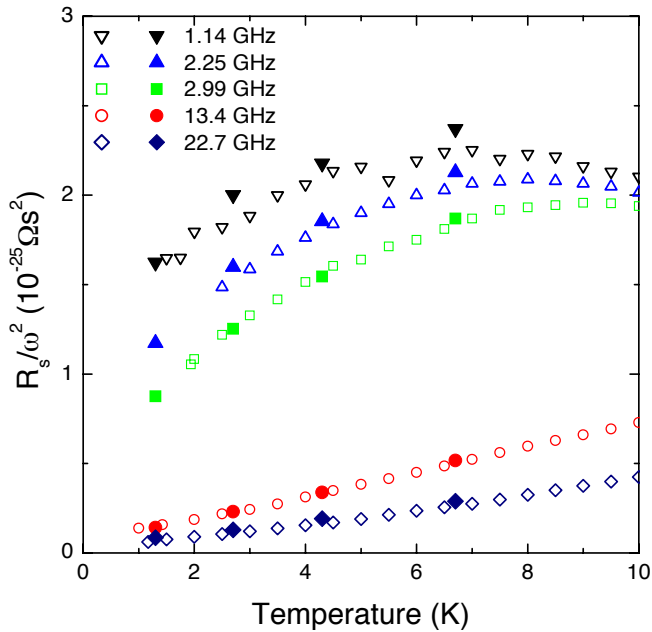}
\caption{\label{fig:figure1} Comparison of surface
resistance measurements made on the same sample of \OrthoII\ using
the broadband experiment (solid symbols) with those from five
microwave resonators (open symbols). The agreement between methods
is excellent.  The data are plotted as $R_s(\omega)/\omega^2$ to
remove the frequency dependence associated with the superfluid
screening.}
\end{figure}

A very convincing verification of the results from multiple sets of
experimental apparatus is provided by a comparison of the broadband
$R_s(\omega,T)$ data with measurements of the {\em same sample} in
five different superconducting resonators.  As seen in
Fig.~\ref{fig:figure1}, the agreement between the
experiments is excellent.  In each of the six experiments the absolute value of
$R_s$ is determined to better than 10\%, with the principle
uncertainty coming from the calibration procedure.

\begin{figure}
\includegraphics[width=\columnwidth]{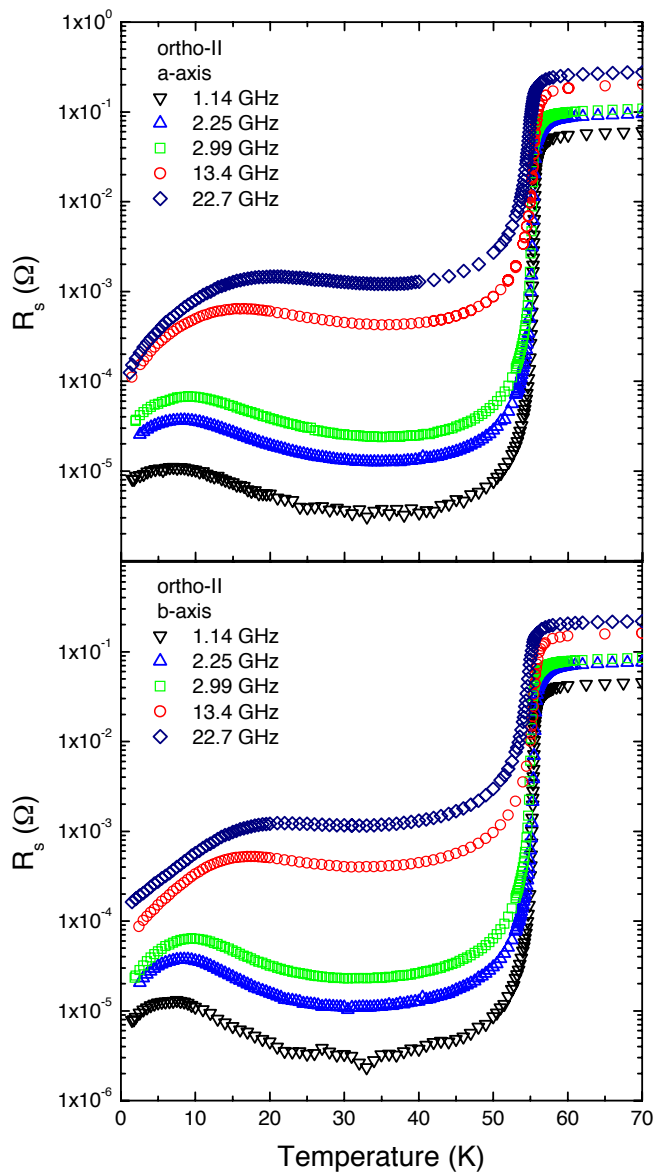}
\caption{\label{fig:figure2} The surface resistance $R_s(T)$ for
Ortho-II ordered \OrthoII\ measured with currents in the
$\hat{a}$-direction (top panel) and $\hat{b}$-direction (bottom panel).}
\end{figure}

\section{Samples}
\label{sec:samples}

The bilayer cuprate compound \YBCO\ has the distinct advantage that
it has a stoichiometric cation composition with a very low level of
cross-substitution, less than one part in $10^4$.  Tunable hole
doping of the CuO$_2$ planes is achieved by varying the oxygen
content of CuO chain layers, which act as a reservoir for oxygen
dopants.  However, the chains add the complication of an extra
one-dimensional conduction channel that acts in parallel with the
two-dimensional channel associated with the \cuplane\ planes.
Varying the oxygen content of \YBCO\ usually occurs at the expense
of introducing off-plane disorder.  Fortunately there are several
highly-ordered phases for the chain oxygen atoms that are stable at
room temperature, and we exploit two of these here.  The first is the fully-oxygenated \Overdoped, with Ortho-I order in which {\em every} CuO chain is nearly completely filled and $T_c=89$~K.  Optimal doping for \YBCO\ occurs at $y=6.93$, corresponding to $T_c=94.0$~K. Increasing the oxygen content beyond 6.93 reduces $T_c$.  Since adding oxygen to the CuO chains is a diffusion-limited process, filling a millimetre-sized crystal beyond $y=6.993$ is impractical. For this publication, we have combined measurements on two different Ortho-I crystals to provide a complete data set: sample~\#1 has $T_c$=89~K and sample~\#2 has $T_c$=91~K. The two samples provided very similar results but with some interesting differences that we attribute to the small difference in oxygen contained in the CuO chains.	

The other ordered phase we have studied is Ortho-II, in which the CuO chains {\em alternate} between full and empty.  Ortho-II order is seen to exist over a wide range of doping, from $y=6.35$ to $y=6.60$, however the most complete chain filling has been found to occur at $y=6.52$.\cite{Ruixing2,Andersen}  The particular sample used for the present microwave study is very close to this, having $y=6.50$ and $T_c=56$~K.

Figure \ref{fig:figure2} presents fixed-frequency microwave absorption measurements for a high quality crystal of \OrthoII\ made in a geometry where the screening currents are along either the $\hat{a}$ or the $\hat{b}$ crystal direction. The surface resistance is seen to drop rapidly, by up to four orders of magnitude, on cooling below
$T_c$, and displays a strong frequency dependence in the microwave
range.  Much of the frequency dependence of $R_s$ is due to the
factor of $\omega^2$ in Eq.~\ref{eqn:RX} that results from screening by the superfluid. The remaining frequency dependence is due to the quasiparticle conductivity $\sigma_1$
itself.  The raw surface impedance data for fully doped \Overdoped\ have 
already been published and are not reproduced here.\cite{Hosseini1, Harris1}

\section{Phenomenology}
\label{sec:phenomenology}

For a superconductor, the conductivity can be expressed as the sum
of two conduction mechanisms acting in parallel: one due to charge
carriers in the superconducting condensate (denoted by $\sigma_{\rm
sf}$) and another due to thermally excited quasiparticles (denoted
by $\sigma_{\rm qp}$):
\begin{equation}
\label{eqn:SigmaSum}
\sigma(\omega,T)=\sigma_{\rm sf}(\omega,T)+\sigma_{\rm qp}(\omega,T).
\end{equation}
The superfluid conductivity can be parameterized using the London phenomenology:\cite{London}
\begin{equation}
\label{eqn:London1}
\sigma_{\rm sf}(\omega,T)=\frac{n_s(T)e^2}{m^*}\left( \pi \delta(\omega)-\frac{\rm i}{\omega}\right),
\end{equation}
where the temperature dependent superfluid spectral weight is
represented by $n_s(T)e^2/m^*$. This is an experimentally
accessible quantity that is related to the London penetration depth
in the following manner:
\begin{equation}
\label{eqn:London2}
\frac{n_s(T)e^2}{m^*}=\frac{1}{\mu_{\rm 0}\lambda_L^2(T)} \, .
\end{equation}

The goal of this work is to explore the phenomenology of the
superconducting state without assuming a microscopic model for the
conductivity of thermally excited \dwave\ quasiparticles.  We extract
$\sigma_{\rm qp}(\omega,T)$  from our measurements of the real and
imaginary components of $Z_s(\omega,T)$ using Eq.~\ref{eqn:Zs}, to
obtain both the spectral lineshape and its temperature dependence.
This process would be straightforward if we could measure both $R_s$
and $X_s$ at each frequency. However, measurements of the frequency
dependence of $X_s$ are very difficult to carry out, for reasons
that are closely related to the difficulty in measuring {\em
absolute} penetration depth at a fixed frequency. To carry out the analysis of our broadband data, we therefore use measurements of $X_s$ at one low frequency only (1~GHz) and extend this to higher frequencies using a self-consistent analysis of the
surface resistance data.  We are helped greatly by the fact that the
frequency dependence of $\lambda$ is very slight, due to the
dominant contribution from the superfluid. The small frequency
dependence of $\lambda$ is due solely to the weak screening
contributions from the thermally excited quasiparticles.  Note that we distinguish between the magnetic penetration depth $\lambda=1/\sqrt{\mu_0 \omega \sigma_2}$ and the London penetration depth $\lambda_L$ that appears in Eq.~\ref{eqn:Phenomenology3}.

\subsection{$\hat{a}$-axis Phenomenology}

\begin{figure}
\includegraphics[width=\columnwidth]{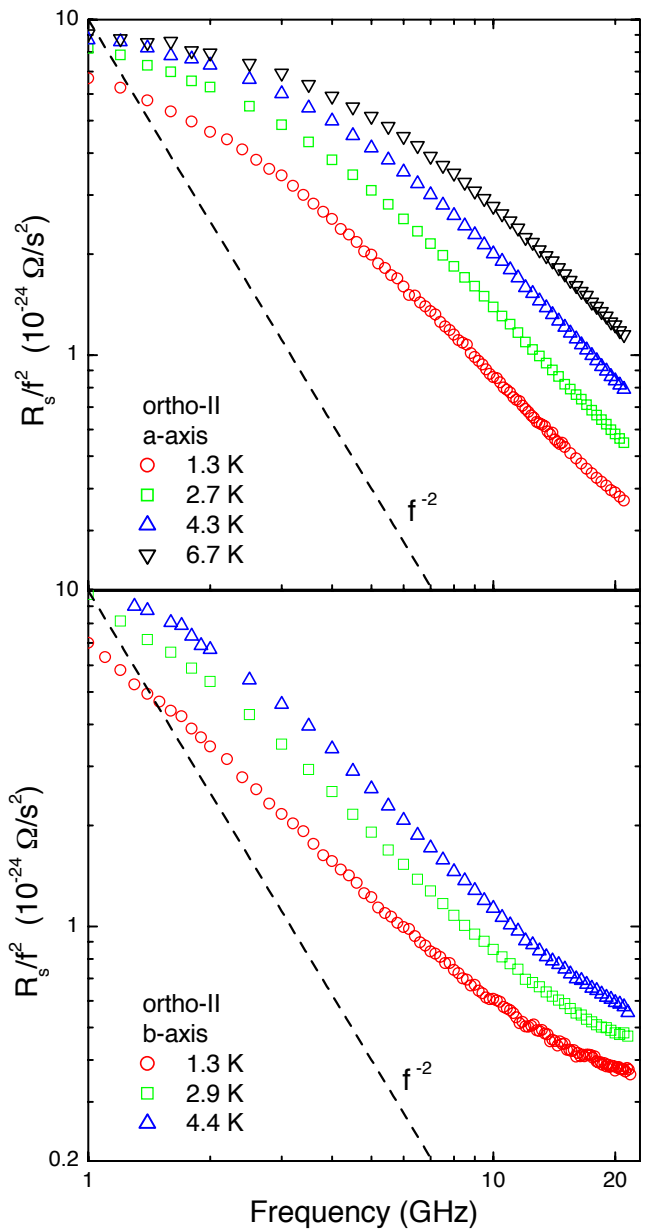}
\caption{\label{fig:figure3} Broadband absorption
measurements of $R_s$ for {\mbox Ortho-II} ordered \OrthoII\ with screening currents
parallel to the crystal $\hat{a}$-axis (top panel) and $\hat{b}$-axis (bottom panel). Plotting the data as $R_s/f^2$ gives a quick means of estimating the conductivity spectrum from the raw data (see Eq.~\ref{eqn:RX}). Here we emphasize that the
frequency dependence does not follow $1/f^2$ as it would for a
Drude conductivity, but rather the power is closer to $1/f^{1.5}$
over most of the range.}
\end{figure}

We begin by analyzing the microwave conductivity for currents parallel to the crystal $\hat{a}$-axis. This is perpendicular to the CuO chains, so these should make a
negligible contribution to the $\hat{a}$-axis conductivity. In
addition, we expect that the \cuplane\ bilayer should lead to two
nearly identical contributions to $\sigma_{ a}(\omega)$, since the
splitting between bonding and antibonding combinations of the planar
wavefunctions is expected to be weak.\cite{Atkinson} We will
therefore model the $\hat{a}$-axis conductivity by a {\em single}
conductivity spectrum due to {\em one} effective two-dimensional
band.

There is currently no accepted phenomenology to describe the
detailed quasiparticle conductivity of a \cuplane\ plane. Several
authors \cite{Berlinsky2, Waldram97, Bonn93} have assumed a Drude-like form:
\begin{equation}
\label{eqn:Drude}
\sigma_{\rm qp}(\omega,T)=\frac{n_n(T)e^2}{m^*}  \frac{1}{ \rm{i} \omega+1/\tau(T)}
\end{equation}
where $n_n(T)e^2/m^*$ is the temperature dependent quasiparticle
spectral weight (often referred to as `normal fluid') and
$1/\tau(T)$ is the thermally-weighted and energy-averaged
quasiparticle transport relaxation rate.  While this particular model has
proven useful for interpreting qualitative features of the
electrodynamics, there is no {\em a priori}
reason why it should model the quasiparticle conductivity correctly.
In fact, broadband measurements demonstrate that the lineshape is
distinctly {\em non}-Drude-like.\cite{TurnerPRL,corson00} A very
clear and simple demonstration of this is obtained by plotting
$R_s/f^2$ at fixed temperature $T$, as shown in figure
Fig.~\ref{fig:figure3} for \OrthoII.  Within the low temperature
approximation of Eq.~\ref{eqn:RX}, $R_s/f^2$ is proportional to
$\sigma_1(\omega)$. The key point is that $R_s/f^2$ does not follow
a $1/f^2$ power law at high frequencies, as would a
Drude spectrum. The data are much closer to following a $1/f^{1.5}$
power law. Consequently, we have proposed a more realistic
phenomenological lineshape to describe the $\hat{a}$-axis
quasiparticle conductivity spectrum in \YBCO:\cite{TurnerPRL}
\begin{equation}
\label{eqn:Phenomenology1}
\sigma_{1a}(\omega,T)=\frac{\sigma_{\rm 0}(T)}{1+[\omega\Lambda(T)]^{y(T)}}
\end{equation}

\noindent where $\sigma_{\rm 0}(T)$ represents the zero frequency
limit of the quasiparticle conductivity, $\Lambda(T)$ the inverse
spectral width, and $y(T)$ the anomalous power.  Note that this is
equivalent to a Drude model if we set $y=2$.  

For the spectrum with $y\neq2$, several conditions apply to the use of Eq.~\ref{eqn:Phenomenology1}.\cite{Ozcan} The expression applies to positive frequencies only, and $y$ must be greater than one or the oscillator-strength sum rule will not be satisfied ({\em i.e.} $\sigma_1(\omega)$ will not be integrable). A more subtle point is that the exponent $y$ must always eventually reach a high frequency value of 2 since the energy-dependent relaxation rate is cut off by the thermal windowing.  However, Eq.~\ref{eqn:Phenomenology1} provides excellent fits to the spectra presented in this article.
Furthermore, Eq.~\ref{eqn:Phenomenology1} is very useful in modeling the
quasiparticle contribution to the imaginary part of the conductivity, because it can be extrapolated and integrated {\em beyond} the frequency range of our measurements.  It will be shown below that success in preserving the oscillator strength sum rule in the
superconducting state supports the use of this phenomenological form
as a means of accounting for the normal fluid contribution to
screening. Since the electrical conductivity is a causal response
function, the real part of the conductivity determines the imaginary
part via a Kramers-Kronig relation:\cite{OrlandoDelin}
\begin{equation}
\label{eqn:Phenomenology2}
\sigma_{2a}(\omega,T)=-\frac{2 \omega}{\pi}{\cal P}\int_0^\infty {\rm d}\omega^{\prime}\frac{\sigma_{1a}(\omega^{\prime},T)}{{\omega^{\prime}}^2-\omega^2} 
\end{equation}
where ${\cal P}$ denotes the principal part of the integral.

\noindent Using this in conjunction with the measured superfluid
response (via Eq.~\ref{eqn:London1}) yields a complete expression
for the imaginary part of $\sigma(\omega,T)$ in Eq.~\ref{eqn:Zs}:
\begin{equation}
\label{eqn:Phenomenology3}
\sigma_2(\omega,T)=\frac{1}{\mu_0\omega\lambda_L^2(T)} +\sigma_{2a}(\omega,T).
\end{equation}

The strategy is to fit broadband $R_s(\omega,T)$ data at a given $T$
using the phenomenological model for the quasiparticle conductivity
(Eqs.~\ref{eqn:Phenomenology1} and \ref{eqn:Phenomenology2}) along
with measurements of the magnetic penetration depth to parameterize
the superfluid conductivity (Eq.~\ref{eqn:London1}). For the
spectrum at each temperature, we obtain best-fit parameters for
$\sigma_{\rm 0}(T)$, $\Lambda(T)$ and $y(T)$.  Finally, we generate
experimental estimates for the real part of the quasiparticle
conductivity by solving Eq.~\ref{eqn:Zs} for $\sigma_{1a}(\omega,T)$
and assuming $\sigma_{2a}(\omega,T)$ as given by the best-fit
parameters.  In many respects this procedure is akin to that used in
infrared spectroscopy studies of superconductors, where one needs to
extrapolate the measured reflectance to zero frequency in order
to utilize Kramers-Kronig relations.\cite{Basov}  The difference is
that we need a model lineshape to effectively extrapolate our
measurements beyond our maximum measurement frequency $f^{\rm max}$, but we do so ultimately for the same purpose.  This technique requires sufficient curvature within the experimental bandwidth ({\em i.e.} $\Lambda^{-1}\!<\! f^{\rm max}$) for a reliable extrapolation.  This approach will be best for low $T$ data, where $\Lambda^{-1}$ will be governed by elastic scattering of \dwave\ quasiparticles from very dilute defects in our high purity samples.  As we turn to modeling the data at higher $T$, inelastic scattering processes ought to push $\Lambda^{-1}$ well beyond $f^{\rm max}$.
Fortunately, as the $\sigma_{1a}(\omega,T)$ spectrum broadens, the
quasiparticle contribution to $\sigma_2(\omega,T)$ becomes
negligible throughout our bandwidth.  In this case, one can set
$\sigma_{2a}(\omega,T)=0$ and extract $\sigma_{1a}(\omega,T)$ from
$R_s(\omega,T)$ and $\lambda_L(T)$ without requiring any of the
corrections discussed above.  
 
\subsection{$\hat{b}$-axis Phenomenology}

Figure \ref{fig:figure3} shows the broadband $\hat{b}$-axis surface
resistance data for \OrthoII\ plotted as $R_s/f^2$ versus $f$. While
the $\hat{a}$-axis data seem to asymptotically follow $1/f^y$ at high frequencies, the $\hat{b}$-axis data show a change in $y$ beyond 10~GHz. In a previous publication we argued that \YBCO\ should have a quasi-one-dimensional band whose low energy
excitations in the superconducting state are primarily chain-like,
and will therefore only contribute to charge conduction along the
$\hat{b}$-axis.\cite{Harris1} The chain scattering rate is
expected to be much larger than that for the \cuplane\ planes, and
should therefore give rise to a very broad contribution to
$\sigma_{1b}(\omega, T)$ that can be approximated as a temperature
dependent constant over the experimental bandwidth in our analysis. The change in $y$ observed above 10~GHz in the $\hat{b}$-axis spectra, but not in the $\hat{a}$-axis, is due to this very broad component. In this case the $\hat{b}$-axis microwave conductivity can then be expressed as a sum of contributions from the two-dimensional bands
(a term identical in form to Eq.~\ref{eqn:Phenomenology1}) plus a
constant attributed to the quasi-one-dimensional band:
\begin{equation}
\label{eqn:Phenomenology4} \sigma_{1b}(\omega, T) = \frac{\sigma_{\rm 0}(T)}{1+[\omega \Lambda
(T)]^{y(T)}} + \sigma_1^{1D} (\omega\rightarrow 0,T) .
\end{equation}
\noindent  In general, the parameters $\sigma_{\rm 0}$, $\Lambda$ and $y$
for $\hat{b}$-axis currents will not be required to match those used to fit $\hat{a}$-axis data. We extract $\sigma_{1b}(\omega, T)$ from $R_s$ data using measurements of $\lambda_L$ and the modeled $\sigma_{2b}(\omega, T)$ as described in the previous section (a constant in
Eq.~\ref{eqn:Phenomenology4} will give no contribution to
$\sigma_{2b}$). The procedure will have limitations similar to
those for the $\hat{a}$-axis case, but the addition of a fourth
parameter $\sigma_1^{1D}$ renders the analysis less reliable as
$\Lambda^{-1}$ approaches the experimental bandwidth at higher $T$.

\section{Quasiparticle Conductivity}
\label{sec:conductivity}

\begin{figure}
\includegraphics[width=\columnwidth]{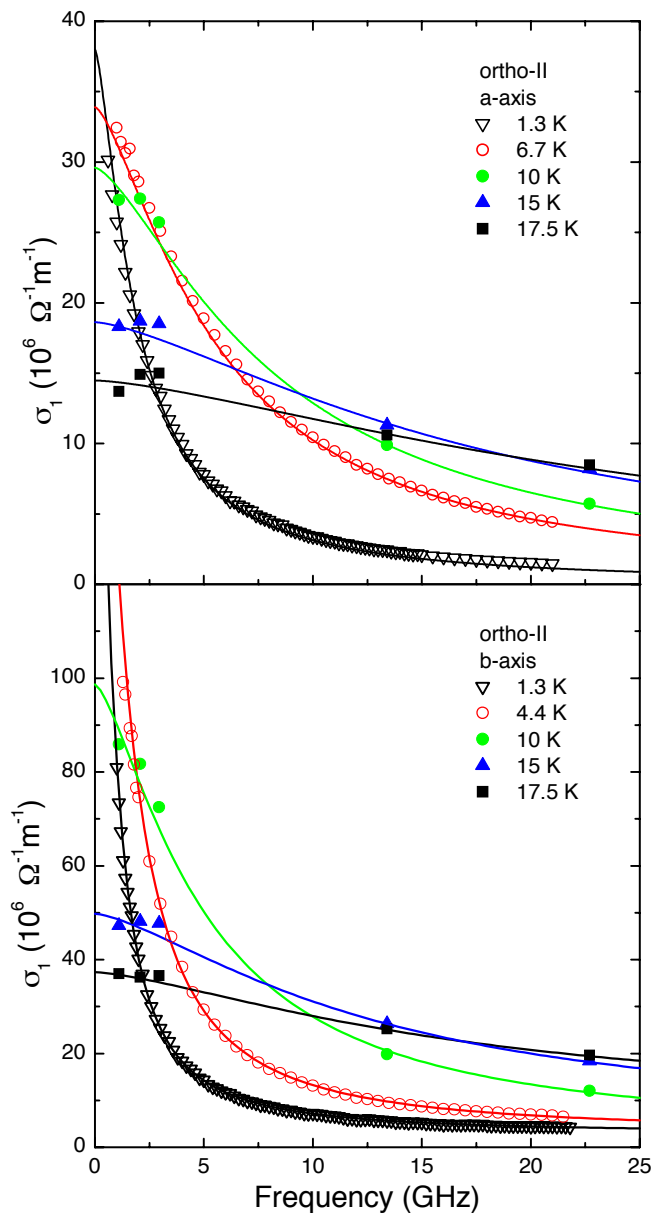}
\caption{\label{fig:figure4} Representative curves of the quasiparticle
conductivity $\sigma_1(\omega,T)$ for Ortho-II ordered  \OrthoII\
measured with currents in the $\hat{a}$-direction (top panel) and $\hat{b}$-direction (bottom panel). The fits shown use the phenomenological model of Eq.~\ref{eqn:Phenomenology1}.  The low temperature cusp-like shape and high frequency power law ($\sim 1/f^{y}$ with $y\sim1.5$) are features expected for a $d$-wave superconductor in the presence of weak-limit impurity scattering, and the spectrum broadens quickly as other scattering mechanisms are introduced at higher temperatures.}
\end{figure}

\begin{figure}
\includegraphics[width=0.99\columnwidth]{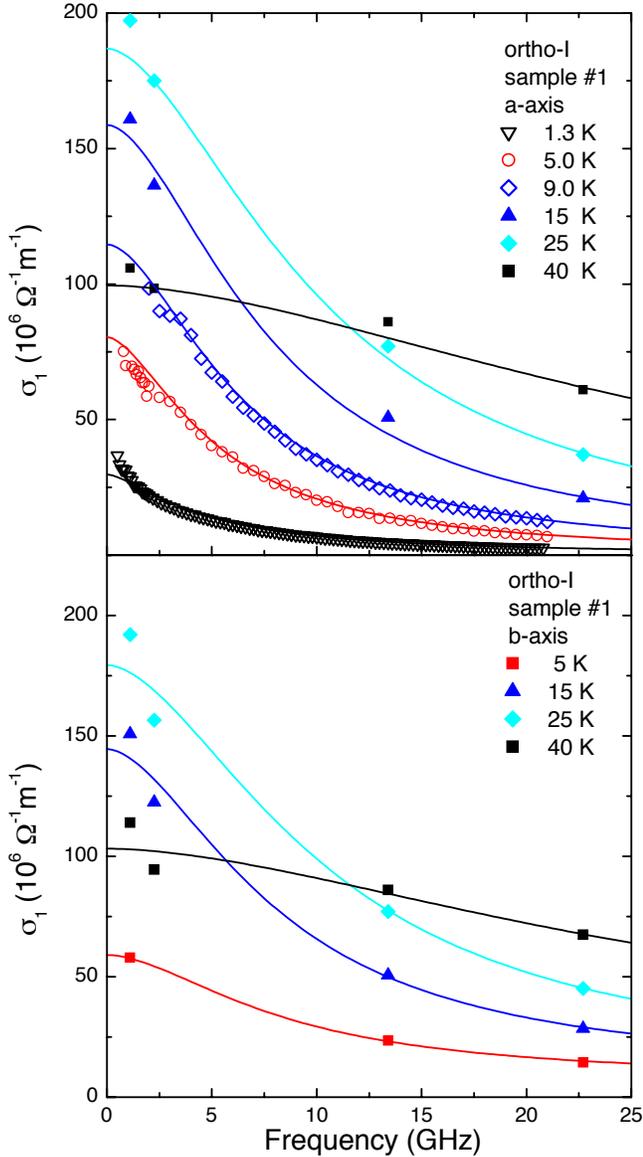}
\caption{\label{fig:figure5} Representative curves of the quasiparticle conductivity $\sigma_1(\omega,T)$ for sample~\#1 of Ortho-I ordered  \Overdoped\ measured with currents in the $\hat{a}$-direction (top panel) and $\hat{b}$-direction (bottom panel). The solid lines are fits to the data with Eq.~\ref{eqn:Phenomenology1} and Eq.~\ref{eqn:Phenomenology4} for the $\hat{a}$ and $\hat{b}$-axis respectively. At the lowest temperature the cusp-like spectrum is evident, however a more Drude-like spectrum is recovered at higher temperatures, consistent with our earlier findings for this Ortho-I ordered \Overdoped.}
\end{figure}

\begin{figure}
\includegraphics[width=\columnwidth]{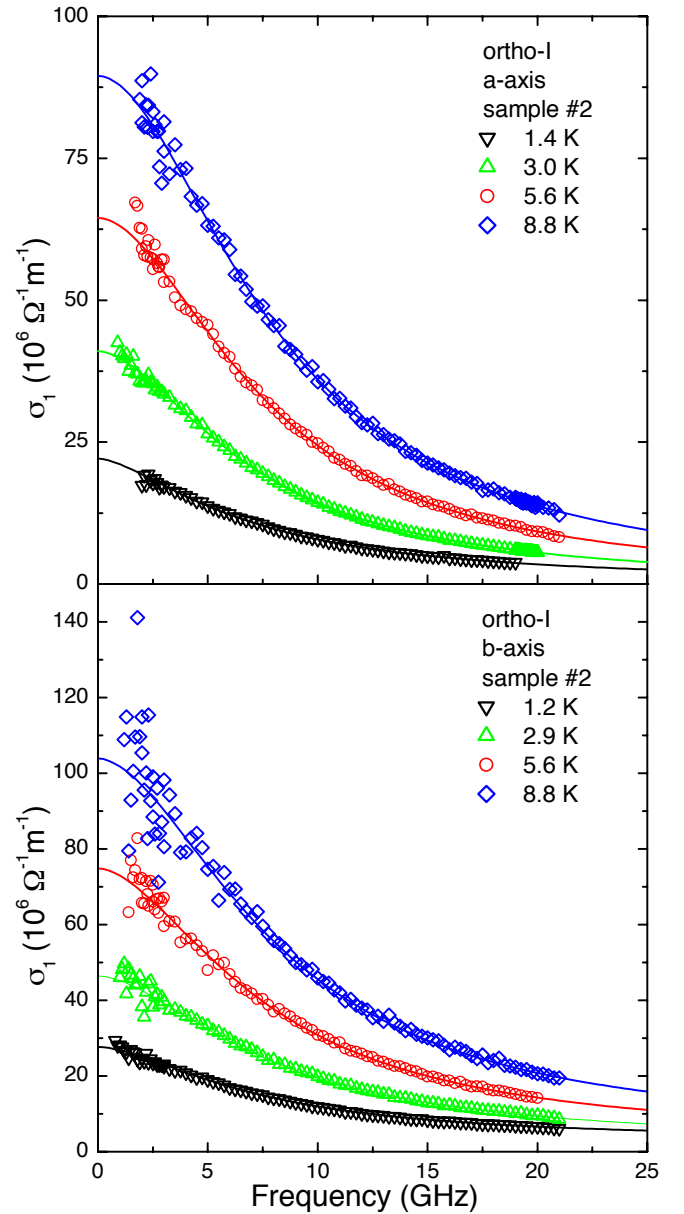}
\caption{\label{fig:figure6} The low temperature quasiparticle
conductivity $\sigma_1(\omega,T)$ for sample~\#2 of Ortho-I ordered  \Overdoped\
measured using the bolometric method with currents in the $\hat{a}$-direction (top panel) and $\hat{b}$-direction (bottom panel).  The spectrum is very similar to that of sample~\#1, although some differences in the fit parameters will be discussed.}
\end{figure}

We show the real part of the quasiparticle conductivity spectrum for
both in-plane principle directions at representative temperatures
for both dopings in Figs.~\ref{fig:figure4}--\ref{fig:figure6},
as obtained from our surface impedance measurements using the
procedures described above.  The figures are comprised of data from the broadband experiment (open symbols) and the microwave resonators (solid symbols).  We include results from two different Ortho-I samples, denoted sample \#1($T_c$=89~K) and sample \#2 ($T_c$=91~K), because we did not have a complete set of data (both crystal directions, all sets of apparatus) for a single sample. As will be shown below, the two samples provided very similar results but with some interesting differences that we attribute to the small difference in oxygen contained in the CuO chains.  In Figs.~\ref{fig:figure4}--\ref{fig:figure6} the smooth lines represent the best fit
lineshapes obtained using Eqs.~\ref{eqn:Phenomenology1} and
\ref{eqn:Phenomenology4} to describe the $\hat{a}$ and
$\hat{b}$-axis results, respectively.  At each temperature our model
lineshape appears to reliably capture the high frequency behaviour,
however our expressions do not accurately follow the data at the
very lowest frequencies, particularly at $T=1.3$~K.  Note that our
model for the two-dimensional conductivity term is necessarily
concave downwards at low frequency while the lowest temperature
bolometry data clearly suggest a cusp. Nonetheless, we argue that
our model lineshape provides a reliable means of extrapolating our
measurements to high frequency and provides a reasonable measure of
the width and spectral weight of the conductivity peaks.

A key point made in the preceding section is that the quasiparticle
conductivity spectra we have measured are distinctly more cusp-like
than the Lorentzian line shape of a Drude model. In
Fig.~\ref{fig:figure8} we present the best-fit values of $y(T)$ that
were used in Eq.~\ref{eqn:Phenomenology1} to describe the
conductivity data extracted from the bolometry data.  Error bars
indicate the standard deviation obtained from fitting to
$R_s(\omega,T)$ as a function of $\omega$ at each given temperature.
Two key features are important to observe: First, $y(T)$ appears to
be crudely $\hat{a}\!:\!\hat{b}$ isotropic for each doping.  Second,
$y(T)$ displays a relatively weak temperature dependence
characterized by a slow increase with increasing temperature, and
appears to saturate at approximately $1.75\pm0.10$ and $1.45\pm0.05$
for the overdoped and underdoped samples, respectively.

\begin{figure}
\includegraphics[width=0.95\columnwidth]{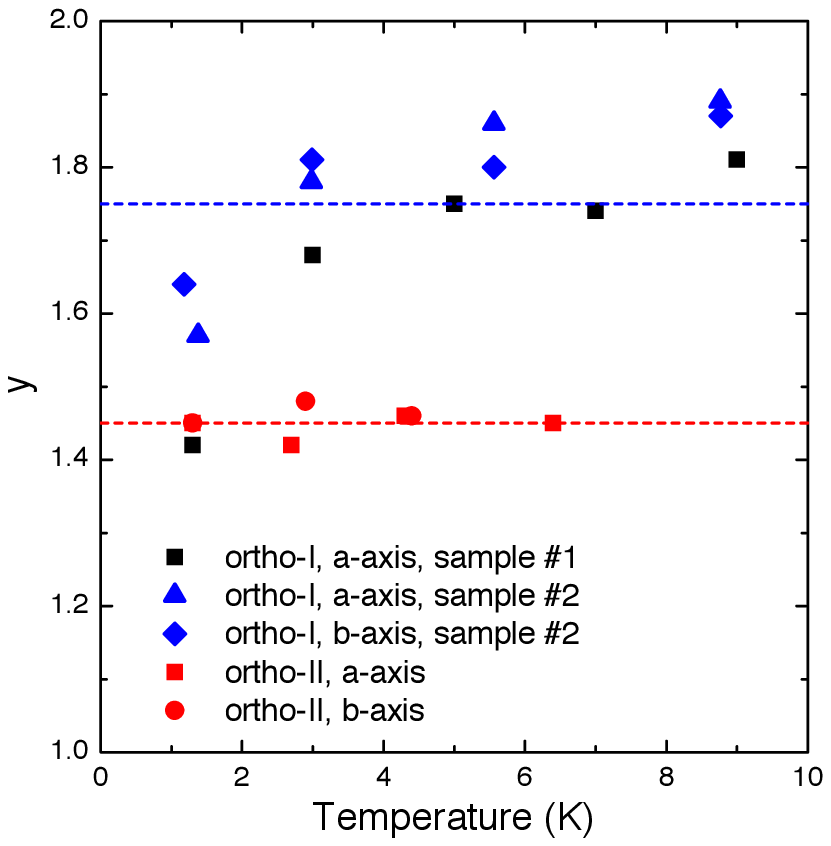}
\caption{\label{fig:figure7} The exponent $y$ of
Eqs.~\ref{eqn:Phenomenology1} and \ref{eqn:Phenomenology4} obtained
from fitting to broadband measurements of the absorption spectrum
for \Overdoped\ and \OrthoII\ samples.  The exponent $y$ describes
the high frequency decay of the quasiparticle conductivity curves
shown in Figs.~\ref{fig:figure4}--\ref{fig:figure6}.  The dashed lines indicate the
asymptotic values of 1.45 (Ortho-II) and 1.75 (sample~\#1, Ortho-I) subsequently
used for fitting to the higher temperature spectra obtained with the
fixed-frequency resonators. Sample~\#2 of the Ortho-I material shows larger values of $y$, corresponding to broader $\sigma_1(\omega)$ curves, which we attribute to a slightly increased level of oxygen disorder in that sample.  Despite this, the value of $y$ is still inconsistent with the Drude value of $y=2$.}
\end{figure}

\begin{figure}
\includegraphics[width=\columnwidth]{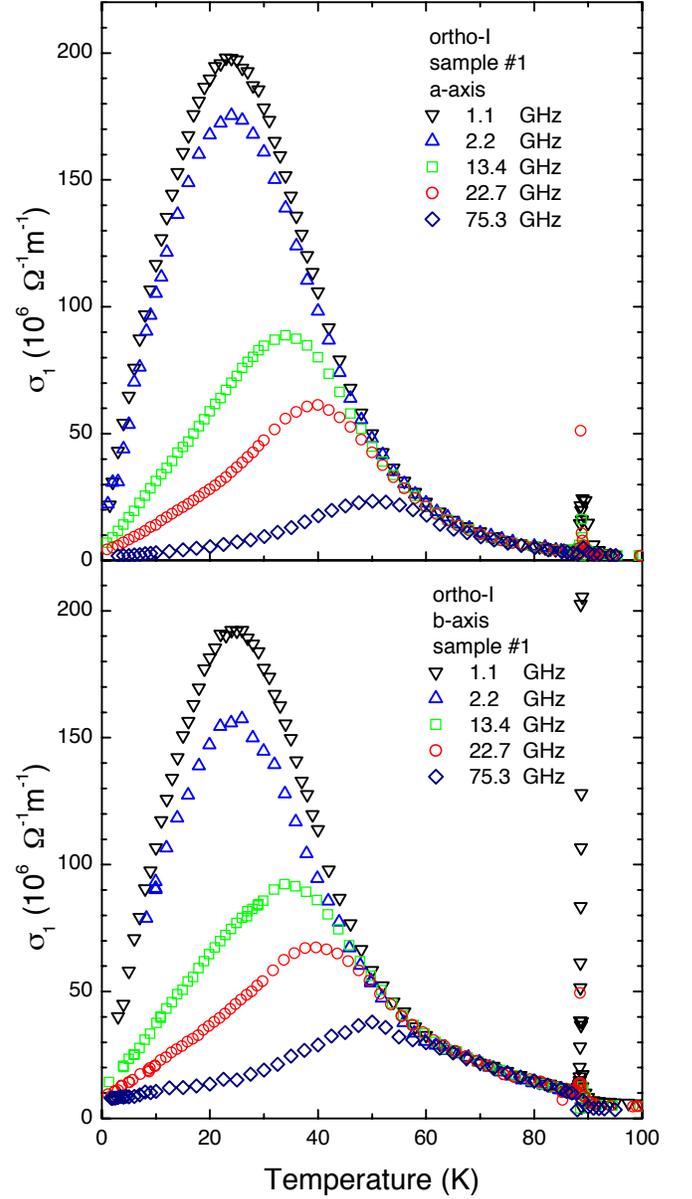}
\caption{\label{fig:figure8} The temperature evolution of the
quasiparticle conductivity $\sigma_1(T)$ for sample~\#1 of overdoped \Overdoped\
measured with currents in the $\hat{a}$-direction (top
panel) and $\hat{b}$-direction (bottom panel). This data has been
processed using the new absolute values of $\lambda(T\to0)$ given in
Table~\ref{table:lambda} resulting in a modification of the absolute
magnitude of the curves from their previously
published\cite{Hosseini1} form.}
\end{figure}

\begin{figure}
\includegraphics[width=\columnwidth]{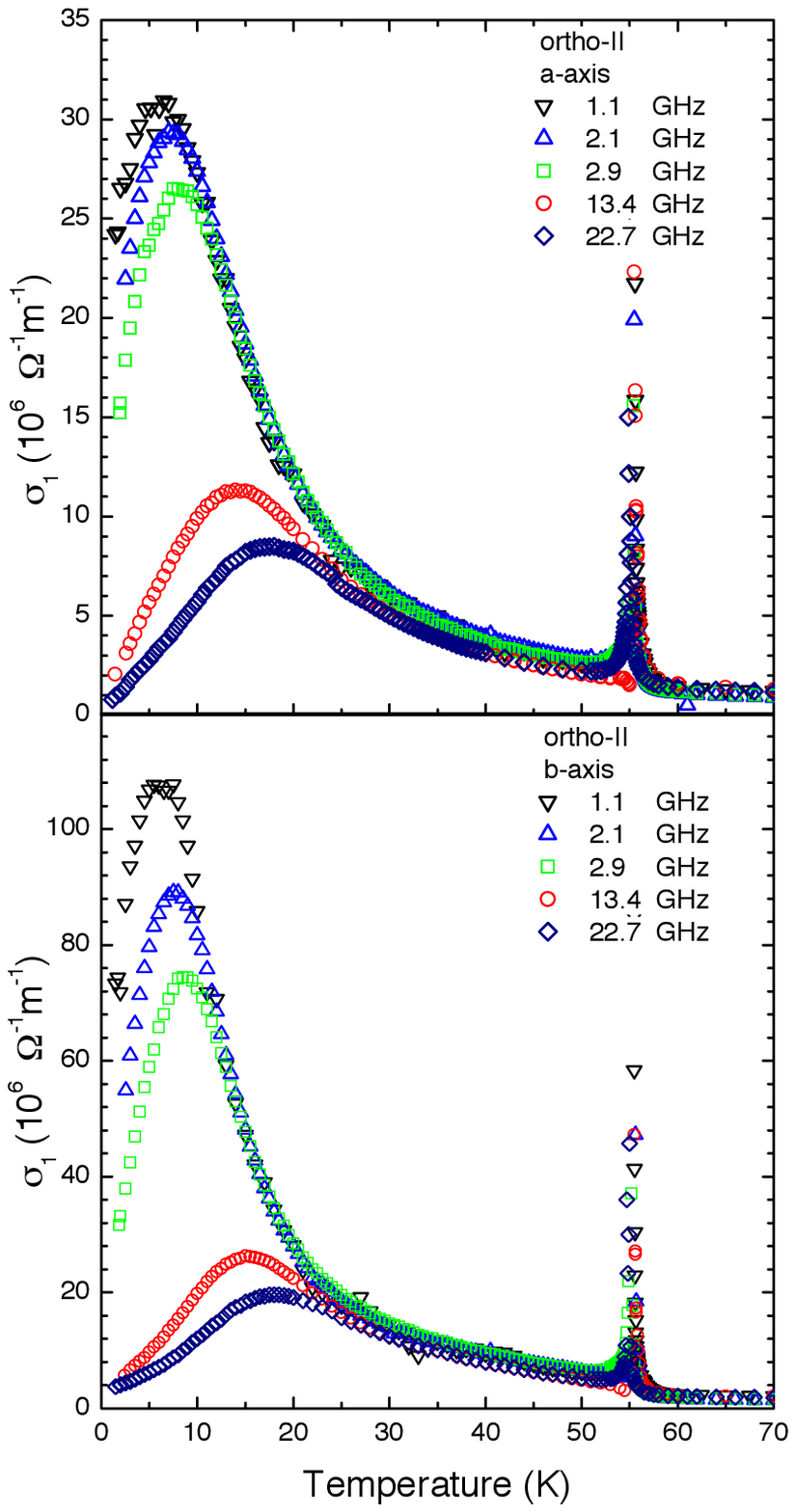}
\caption{\label{fig:figure9} The temperature evolution of the
quasiparticle conductivity $\sigma_1(T)$ for Ortho-II ordered
\OrthoII\ measured with currents in the $\hat{a}$-direction
(top panel) and $\hat{b}$-direction (bottom panel).}
\end{figure}

We now turn to the problem of extracting the real part of the
quasiparticle conductivity from our cavity perturbation experiments.
In this case we only have measurements at discrete frequencies, from
independently calibrated experiments. To help constrain the fits
when using such a sparse set of frequencies, we choose to fix $y$ at
the asymptotic high-temperature values of Fig.~\ref{fig:figure8}
before proceeding to extract the real part of the microwave
conductivity from $R_s(\omega ,T)$. We note that changing $y$
ultimately leads to negligible changes in the extracted values of
$\sigma_1(\omega,T)$ within our experimental bandwidth. However, we
will show below that the choice of $y$ has implications for the
integrated quasiparticle spectral weight and that the values
obtained are consistent with the oscillator strength sum
rule.\cite{Tinkham} Representative spectra from the resonator
measurements with fits are included in Figs.~\ref{fig:figure4}--\ref{fig:figure6}.

We show $\sigma_1(\omega,T)$ extracted from all of the cavity
perturbation measurements in Figs.~\ref{fig:figure8} and
\ref{fig:figure9}. Note in each plot there is a  sharp peak in the
curves at $T_c$ due to superconducting fluctuations,\cite{Kamal3DXY,
KamalHP} in addition to the large, broad peak which shifts to higher
temperatures as the measurement frequency is increased.   
These results can be interpreted in the same qualitative manner as in
previous work.\cite{Hosseini1,Harris1}  Neglecting the sharp
peaks at $T_c$, $\sigma_1(\omega,T)$ increases smoothly on cooling from the normal state into the superconducting state.  The large increase in $\sigma_1(\omega,T)$ below $T_c$ can be attributed to a rapid decline in the quasiparticle transport relaxation rate below $T_c$, combined with an increase in the condensate fraction.\cite{Bonn92,Bonn93} The frequency dependence of $\sigma_1(\omega,T)$ becomes apparent at our measurement frequencies for temperatures below 55~K for the Ortho-I sample and 24~K for the Ortho-II sample. This is in accord with the idea that the peak in the temperature
dependence comes from the development of long quasiparticle
transport lifetimes.

\begin{table}
\begin{tabular}{|c|c|c|c|c|c|c|}
\hline ~Phase~ &~$T_c$(K)~& ~6+y~ & ~~$\lambda_a$(nm)~~
& ~~$\lambda_b$(nm)~~ & ~~~$\lambda_c$(nm)~~~  \\
\hline Ortho-I & 89&6.995 & $103\pm 8$ & $80\pm 5$&  $635\pm 50$\\
\hline
Ortho-II &56& 6.50  &$202\pm 22$ & $140\pm 28$ &  $7500 \pm 480$ \\
\hline
\end{tabular}
\caption[Experimental values of the anisotropic magnetic penetration
depth.] {Experimental values of the anisotropic magnetic penetration
depth measured by a zero field ESR technique.\cite{TamiPRB} The
absolute value of $\lambda$ sets the absolute value of the
conductivity.} \label{table:lambda}
\end{table}

The degree to which the extraction of $\sigma_1(\omega)$ from
$R_s(\omega)$ is affected by uncertainties in $\lambda(T \to 0)$ is
an important issue that must be addressed.  From Eq.~\ref{eqn:RX},
it is clear that $\sigma_1 \approx R_s/\lambda^3$.  A 20\% uncertainty in
the absolute value of $\lambda$ therefore produces almost a factor of
two error in the conductivity although it does not affect qualitative
features in $\sigma_1(\omega,T)$.  In previous work, we have taken
our $\lambda(T\to0)$ values from the literature, usually from
infrared reflectivity experiments,\cite{Basov} which are able to
resolve the $\hat{a}$--$\hat{b}$ plane anisotropy. Very recently we
have developed an accurate means of determining
$\lambda_L(T\rightarrow 0)$ from broadband zero-field ESR absorption
spectroscopy of Gd$^{3+}$ ions in \GYBCO.\cite{TamiPRB}  The low
temperature limiting values of the London penetration depth are
quoted in Table \ref{table:lambda} for samples of the same quality
as those studied in this article, and they differ significantly from
the values used previously.\cite{Hosseini1,Harris1} Note that we
quote our values of $\lambda_L(T\rightarrow 0)$ in
Table~\ref{table:lambda} to better than 10\% accuracy based on the
redundant measurements and self-consistency checks that were built
into our analysis protocol for that work.\cite{TamiPRB} We have
reprocessed the previously published\cite{Hosseini1,Harris1}
$\hat{a}$ and $\hat{b}$-axis conductivity data for \Overdoped\ and
graphed the updated $\sigma_1(\omega,T)$ data in
Fig.~\ref{fig:figure8}.

\subsection{Spectral Width}
\label{subsec:spectralwidth}

\begin{figure}
\includegraphics[width=\columnwidth]{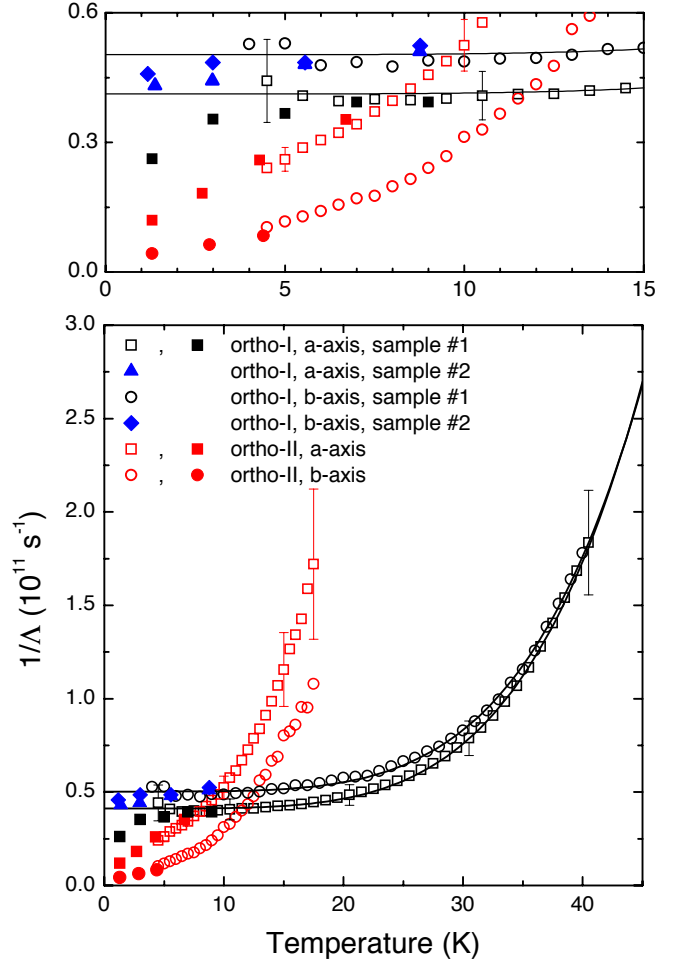}
\caption{\label{fig:figure10} The spectral width parameter
$1/\Lambda(T)$ for both in-plane orientations and both oxygen
dopings obtained from modeling the quasiparticle conductivity
spectra shown in Figs.~\ref{fig:figure4}--\ref{fig:figure6} with Eqs.~\ref{eqn:Phenomenology1} and \ref{eqn:Phenomenology4}. The open symbols are from fits to the cavity perturbation data and the solid symbols are from fits to
the bolometry data.  The upper panel details the low temperature behaviour.}
\end{figure}

Having extracted the real part of the quasiparticle conductivity
from all of our surface impedance measurements, we now turn to a
detailed discussion of the best-fit parameters arising from the
model in Eqs.~\ref{eqn:Phenomenology1} and \ref{eqn:Phenomenology4} for
the $\hat{a}$-axis and $\hat{b}$-axis data respectively.  To begin,
we show the spectral width parameter $\Lambda^{-1}(T)$ for both
in-plane orientations and both oxygen dopings in
Fig.~\ref{fig:figure10}.  Representative error bars have been
included at select temperatures and indicate the standard deviation
from fitting to $R_s(\omega,T)$ spectra as a function of $\omega$ at
each given $T$.  The phenomenological parameter
$\Lambda^{-1}(T)$ is a measure of the energy-averaged
quasiparticle transport relaxation rate, even though the energetics of the
\dwave\ superconducting state enter in a nontrivial manner.\cite{HPS1, HPS2, HarrisPhD, schachinger02} Ultimately, a full microscopic treatment of the charge transport should be used to fit the details of $\sigma_1(\omega , T)$. In the meantime, the phenomenological
parameters provide useful insight.

Upon examining the data for the Ortho-I sample in
Fig.~\ref{fig:figure10}, four key conclusions can be drawn. First,
$1/\Lambda(T)$ appears to be $\hat{a}$:$\hat{b}$ {\em isotropic} to
within our experimental uncertainties.  Second, fitting to the
parameters arising from the resonator data indicates an average for
the low temperature limit of $\overline{\Lambda^{-1}} \sim (45.8\pm
1.0) \times 10^9~{\rm s}^{-1}$ and a temperature dependent component
that grows rapidly, either as $T^x$ with $\overline{x}=4.6\pm 0.1$,
as indicated by the solid lines in the plot, or even exponentially.
These observations are consistent with those made in our previous
publications.\cite{Hosseini1,Harris1,TurnerPRL}  Third, we note that
the absolute values of the damping $1/\Lambda(T)$ reported here
(with $y=1.75$) are essentially identical to the values of the Drude
scattering rate $\tau^{-1}(T)$ (with $y=2$) reported in our previous
work. We can therefore regard the parameter $\Lambda$ in our
phenomenological expression for the conductivity as a meaningful
measure of the average quasiparticle transport lifetime.  The fourth and final point is that sample~\#1, with more complete oxygen filling, reveals lower $1/\Lambda$ values, while the increased number of oxygen vacancies in sample~\#2 seem to cause a small broadening of the $\sigma_1(\omega)$ spectra.  However, although the spectra are broader, the approximate isotropy of $\Lambda$ is retained.

Turning now to the Ortho-II results presented in
Fig.~\ref{fig:figure9}, a quite different scenario is observed.  First, the spectral width parameter is evidently $\hat{a}$:$\hat{b}$ {\it anisotropic}. Furthermore,
$1/\Lambda(T)$ remains strongly temperature-dependent all the way
down to 1.2~K. This latter observation was a critical piece of
evidence that led to the conclusion that the $\hat{a}$-axis
quasiparticle conductivity of Ortho-II ordered \OrthoII\ is
dominated by relatively weak (or Born-limit) scattering from static
crystalline defects at low temperature.\cite{TurnerPRL,HPS1} On the
other hand, the linear-in-temperature regime apparently extrapolates
back to a nonzero value of $1/\Lambda(T\!=\!0)$, which is not in
agreement with simple pictures of elastic scattering from weak
point-like Coulomb potentials.  Rather than scaling with $T$, we shoed that the width of the quasiparticle conductivityspectrum scales as $T+T_{\rm 0}$ with $T_{\rm 0}=2.0$~K.\cite{TurnerPRL}

The contrast between the overdoped and underdoped values of $1/\Lambda(T)$ is striking. As stated above, the spectral width gives the thermally-averaged quasiparticle relaxation rate.  Given that the chemical purity and crystallinity of all
of the samples studied here are nominally the same, it is somewhat
surprising that the conductivity spectra for the two different
dopings evolve in a very different manner.  In fact, this
observation suggests that oxygen doping in the CuO chain layer of
\YBCO\ has significant influence on the scattering of \dwave\
quasiparticles in the adjacent \cuplane\ planes.  In addition, the
large $\hat{a}$:$\hat{b}$ anisotropy of the relaxation rate in {\mbox Ortho-II} is a surprising contrast to the approximate istotropy of {\mbox Ortho-I}. The anisotropy is puzzling since the low energy excitations reside near the four $d$-wave nodes along the diagonals of the Brillouin zone, and these excitations must carry both the $\hat{a}$ and $\hat{b}$-axis currents.

\subsection{Spectral Weight}
\label{subsec:spectralweight}

Our phenomenological fits to the conductivity spectra yield the
parameters $\sigma_{\rm 0}(T)$, $\Lambda(T)$ and $y(T)$.  Rather
than plot $\sigma_{\rm 0}(T)$, the zero-frequency intercept of the
conductivity, it is instead more insightful to present the
integrated spectral weight of the quasiparticle conductivity.  This
quantity is often referred to as the normal fluid oscillator
strength.  We obtain the spectral weight by integrating the fitted
phenomenological form associated with the two-dimensional bands, and
denote this quantity by  $(n_n(T) e^2/m^*)_{2D}$.

The phenomenological spectrum in Eq.~\ref{eqn:Phenomenology1} can be integrated in closed form to give:
\begin{eqnarray}
\label{eqn:SpectralWeight1}
\left(\frac{n_n (T)e^2}{m^*}\right)_{\!2D}& =& \frac{2}{\pi}\int \! d\Omega
\,\sigma_1^{2D}(\Omega, T) \\ \nonumber &=& \frac{2 ~ \sigma_{\rm 0}(T)}{y(T)\Lambda(T)\sin\big(\pi/y(T)\big)}.
\end{eqnarray}

\noindent Values for $y(T)$ are given in Fig.~\ref{fig:figure8} for the
fits to bolometric measurements and values of $y=1.75$ (sample~\#1, Ortho-I) and
$y=1.45$ (Ortho-II) are used for the fits to the cavity perturbation
measurements. We use the values of $\Lambda(T)$ plotted in
Fig.~\ref{fig:figure9}.  By plotting the spectral weight we carry out
an important check on both the experiments and the subsequent
analysis of the data.  The Ferrel-Tinkham-Glover sum
rule\cite{Tinkham} states that the quasiparticle spectral weight
and the superfluid spectral weight must add to a temperature-independent constant $ne^2/m^*$:
\begin{equation}
\label{eqn:SpectralWeight2}
\sum_{\mbox{\tiny \rm Bands}~i} \! \left[\left(\frac{n_n(T) e^2}{m^*}\right)_{\!i} \! +\! \left(\frac{n_s(T) e^2}{m^*}\right)_{\!i} ~\right] =\! \sum_{\mbox{\tiny \rm Bands}~i} \! \left(\frac{n e^2}{m^*}\right)_{\!i} .
\end{equation}

\noindent Here, the London penetration depth of Eq.~\ref{eqn:London2} is related to the superfluid spectral weight $n_s(T)e^2/m^*$ defined as the sum over all bands that participate in superconductivity:
\begin{equation}
\label{eqn:SpectralWeight3}
\frac{n_s(T)e^2}{m^*}\equiv\sum_{\mbox{\tiny \rm Bands }~i}\left(\frac{n_s(T) e^2}{m^*}\right)_{\!i}.
\end{equation}

It should be possible to observe the transfer of spectral weight
between the quasiparticles and the superconducting condensate as a
function  of temperature.  The sum rule does not require complete condensation ---
it is entirely possible that residual quasiparticle oscillator
strength will remain as $T\!\!\rightarrow\!\!0$. We will demonstrate
that in our \YBCO\ samples, up to 3\% of
the low frequency spectral weight fails to condense.  A similar conclusion
has been drawn from microwave measurements on \TBCO\ and \BSCCO,
where at least 10\% of the oscillator strength remains as
$T\!\!\rightarrow\!\!0$.\cite{Ozcan}

\begin{figure}
\includegraphics[width=\columnwidth]{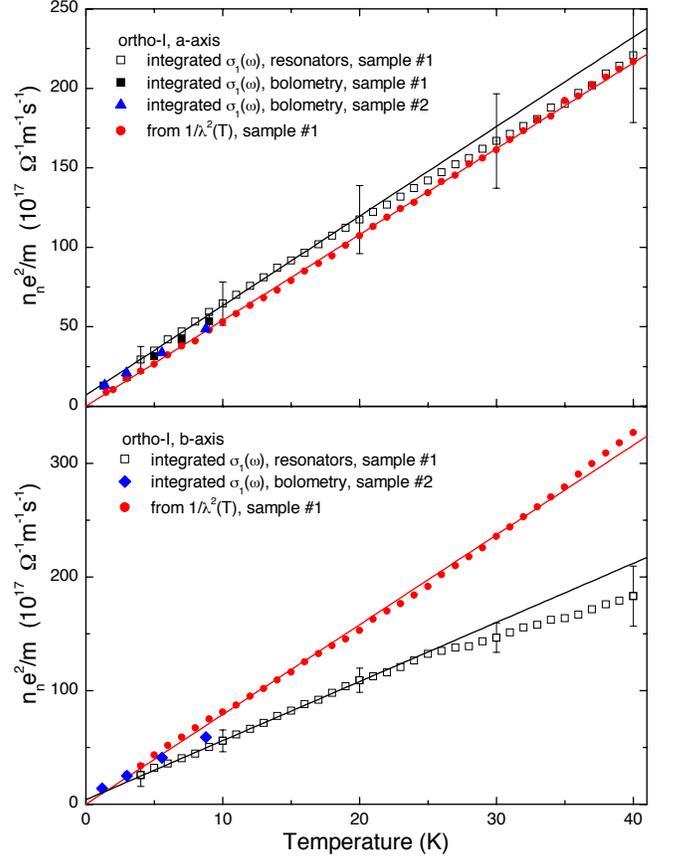}
\caption{\label{fig:figure11} Normal fluid oscillator strength for fully doped \Overdoped\ associated with the two-dimensional CuO$_2$ planar bands.  Measurements of the real and imaginary parts of the conductivity allow this to be determined by two methods: from integrating the $\sigma_1(\omega)$ data (see Eq.~\ref{eqn:Phenomenology1} and Fig.~\ref{fig:figure5}), and from the disappearance of oscillator strength in the superfluid response, as determined by measurements of $\Delta \lambda(T)$ at 1.1~GHz (see Eq.~\ref{eqn:SpectralWeight4}).  The close agreement in slopes for the $\hat{a}$-axis indicates that the conductivity sum rule is obeyed.  This is not the case for the $\hat{b}$-axis where a considerable fraction of the total spectral weight resides in the broad spectral feature in $ \sigma_1(\omega)$ associated with the one-dimensional CuO chains.  The difference in the $\hat{b}$-axis case between sample~\#1 and sample~\#2 indicates that this parameter is very sensitive to oxygen content.}
\end{figure}

In Figs.~\ref{fig:figure10} and \ref{fig:figure11} we show the integrated quasiparticle
spectral weight from the two-dimensional bands of \YBCO.
Representative error bars are shown at selected temperatures and were
calculated from the standard deviation of the best fit parameters.
In addition, we show the corresponding loss of superfluid spectral
weight, defined to be:
\begin{equation}
\label{eqn:SpectralWeight4}
\frac{n_s(T\!=\!0)e^2}{m^*} - \frac{n_s(T)e^2}{m^*} = \frac{1}{\mu_0 \lambda_L^2(T\!=\!0)} - \frac{1}{\mu_0
\lambda_L^2(T)},
\end{equation}

\noindent for  both dopings and for $\hat a$ and $\hat b$ directions.

Errors in the choice of $\lambda_L(T\!\!\rightarrow\!\!0)$ will have
a strong effect on the absolute values of spectral weight obtained
from our analysis.  However, by fortunate coincidence, the
uncertainty in $\lambda_L(T\!\!\rightarrow\!\!0)$ affects the the
quasiparticle oscillator strength and the superfluid spectral weight
in the same way. This can be seen by first noting that at low
temperature and low frequency $\sigma_1\propto
R_s/\lambda_L^3\approx R_s/\lambda_L^3(T\!\!=\!\!0)$.  Second, at
low temperature one can write the following approximation for
Eq.~\ref{eqn:SpectralWeight4}:
\begin{equation}
\label{eqn:SpectalWeight5}
\frac{n_s(T\!=\!0)e^2}{m^*} - \frac{n_s(T)e^2}{m^*} \approx \frac{2\Delta\lambda_L(T)}{\mu_0 \lambda_L^3(T\!=\!0)},
\end{equation}
\noindent where
$\Delta\lambda_L(T)\equiv\lambda_L(T)-\lambda_L(T\!\!=\!\!0)$.
Therefore the change with temperature of the integrated
quasiparticle spectral weight and the superfluid spectral weight
both scale as $1/\lambda_L^3(T\!\!=\!\!0)$.  This is an important result
because it allows us to test our data and analysis for consistency
with the oscillator strength sum rule in a manner that is
insensitive to uncertainties in $\lambda_L(T\!=\!0)$.

\begin{figure}
\includegraphics[width=\columnwidth]{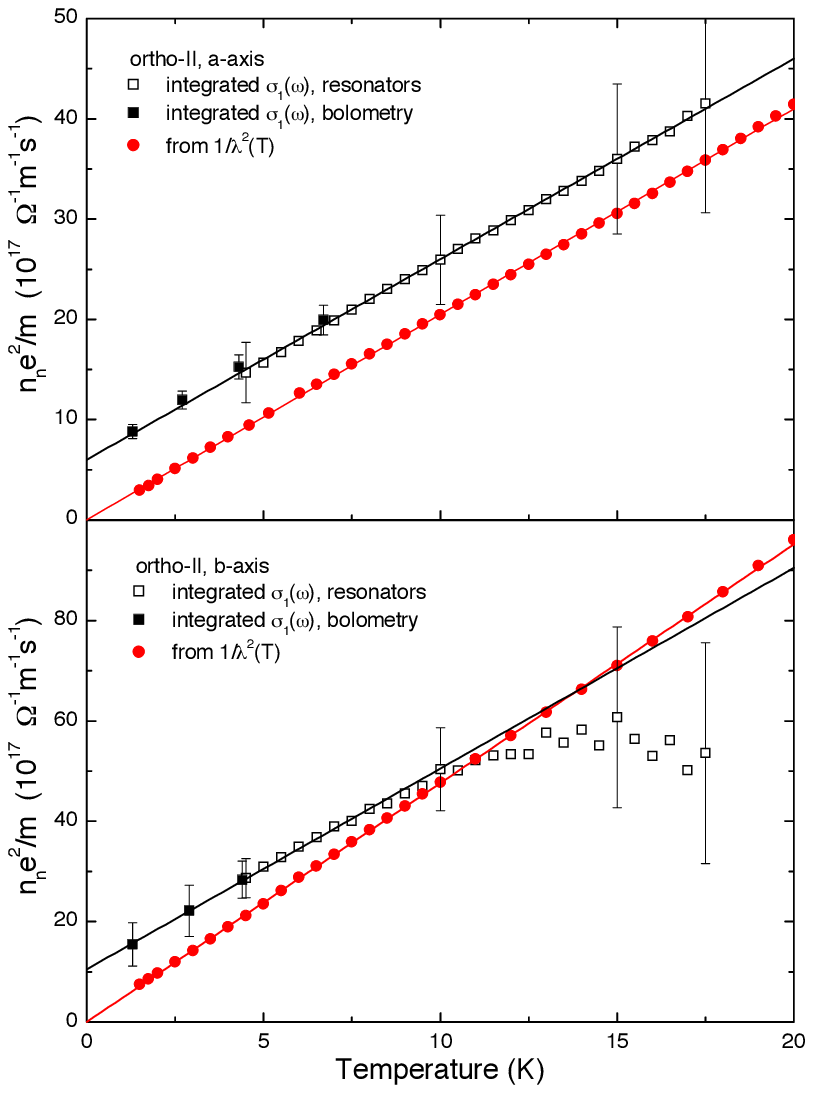}
\caption{\label{fig:figure12} Normal fluid oscillator strength for Ortho-II ordered \OrthoII\ associated with the two-dimensional CuO$_2$ planar bands.  As for the fully doped case, the $\hat{a}$-axis data closely obey the sum rule and the $\hat{b}$-axis data are more complicated.  Note that only integration of $\sigma_1(\omega)$ gives the correct $T\to 0$ offset since our calculation of $n_ne^2/m^*$ from $1/\lambda^2$ assumes there is no residual quasiparticle oscillator strength. In all cases we observe a fraction of the total spectral weight that remains uncondensed (see Table~\ref{table:SpectralWeightIntercepts}) similar to observations in other cuprates.}
\end{figure}

There are three features common to all panels in Figs.~\ref{fig:figure11} and
\ref{fig:figure12}.  First, the superfluid spectral weight determined from measurements of $\Delta \lambda(T)$ varies linearly with temperature at low $T$.  This is a general consequence of having line nodes in the superconducting order parameter.\cite{Scalapino}  Second, the quasiparticle spectral weight attributed to the two-dimensional
bands also varies linearly with temperature at low $T$.  The low $T$ slopes were obtained by linear fitting to the data of  Figs.~\ref{fig:figure11} and
\ref{fig:figure12}, and the results are tabulated in
Table~\ref{table:SpectralWeightSlopes}.  Third, all of the
integrated quasiparticle spectral weight data extrapolate to
nonzero values at $T\!=\!0$ indicating that a finite fraction of the spectral weight does not condense.  Our estimates of the $T\!=\!0$ intercepts are tabulated in
Table~\ref{table:SpectralWeightIntercepts}.  Approximately 0.5\% and 3\% of the total spectral weight remains uncondensed for the {\mbox Ortho-I} \Overdoped\ and {\mbox Ortho-II} \OrthoII\ material respectively.  The large error bars for the $\hat{b}$-axis
results in Tables~\ref{table:SpectralWeightSlopes} and
\ref{table:SpectralWeightIntercepts} are due to the
uncertainty in the fit parameter $\sigma_1^{1D}(\omega\rightarrow
0,T)$ used to capture contributions from the CuO chain electrons.

\begin{table}
\begin{tabular}{|c|c|c|c|}
\hline Doping & Orientation & Spectral Weight & Slope  \\
  &   &   & ($10^{17}\,(\Omega {\rm m s K})^{-1}$)\\
\hline\hline
Ortho-I & $\hat{a}$ & 2D-QP & $5.4\pm 0.7$ \\
sample \#1& $\hat{a}$ & from $1/\lambda^2(T)$ & $5.4$ \\
& $\hat{b}$ & 2D-QP & $5.2\pm 0.7$ \\
& $\hat{b}$ & from $1/\lambda^2(T)$ & $7.9$ \\ \hline
Ortho-I & $\hat{a}$ & 2D-QP & $4.8\pm 0.2$\\
sample \#2& $\hat{b}$ & 2D-QP & $5.9\pm 0.1$ \\ \hline
Ortho-II & $\hat{a}$ & 2D-QP & $2.0\pm 0.5$ \\
& $\hat{a}$ & from $1/\lambda^2(T)$ & $2.0$ \\
& $\hat{b}$ & 2D-QP & $4.0\pm 1.8$ \\
& $\hat{b}$ & from $1/\lambda^2(T)$ & $4.8$ \\ \hline
\end{tabular}
\caption[]
{Temperature slopes of the spectral weight $n_n(T)e^2/m^*$ from the data shown in Figs.~\ref{fig:figure10} and \ref{fig:figure11}. The two-dimensional quasiparticle (2D-QP) slopes given in this table are an average of the slopes of the low temperature resonator and bolometry data in the cases where both measurements exist.}
\label{table:SpectralWeightSlopes}
\end{table}

The results in Table \ref{table:SpectralWeightSlopes} lead to a
critical conclusion: along the $\hat{a}$-axis, for both
dopings, the rate at which the superfluid spectral weight decreases
with temperature is equal in magnitude to the rate at which
quasiparticle spectral weight increases with temperature. Our results
are therefore consistent with the Ferrel-Tinkham-Glover sum rule in
both Ortho-I \Overdoped\ and Ortho-II \OrthoII.  This gives considerable support for the general correctness of the phenomenological lineshape of
Eq.~\ref{eqn:Phenomenology1}. Again, we emphasize that this
agreement is independent of the exact choice of
$\lambda_L^a(T\rightarrow 0)$.

\begin{figure}
\includegraphics[width=\columnwidth]{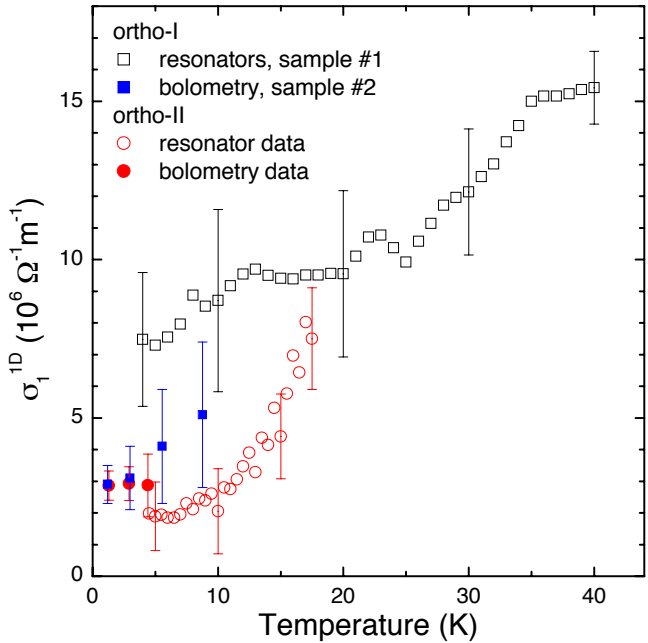}
\caption{\label{fig:figure13} Temperature dependence of the broad spectral feature in the $\hat{b}$-axis conductivity associated with the quasi-one-dimensional CuO chain bands for both Ortho-I and Ortho-II \YBCO.  The conductivity component $\sigma_1^{1D}(\omega\rightarrow 0,T)$ in Eq.~\ref{eqn:Phenomenology4} is frequency independent in our phenomenological model since it is known to possess a width much greater than our highest measurement frequency. }
\end{figure}

The $\hat{b}$-axis results in Table~\ref{table:SpectralWeightSlopes} reveal that the
rate at which the quasiparticle spectral weight attributed to the
two-dimensional bands increases with temperature is {\it less} than
the rate at which superfluid spectral weight decreases
with temperature for both dopings.  This observation is consistent
with the idea that part of the quasiparticle spectral weight resides in the broad spectral feature associated with the quasi-one-dimensional chain-like band.  Previously\cite{Harris1} we have shown that this is a very broad feature and therefore we cannot estimate its spectral weight. The fit parameter $\sigma_1^{1D}(\omega\rightarrow 0,T)$ for both dopings is plotted in Fig.~\ref{fig:figure13}.  Error bars are shown at selected temperatures and indicate one standard deviation in this fit parameter.  The resonator measurements on \mbox{Ortho-I} suggest that $\sigma_1^{1D}(\omega\rightarrow 0,T)$ varies roughly linearly with $T$ and extrapolates to a $T\!=\!0$ intercept of
$(7\pm3)\times10^6\,\Omega^{-1} {\rm m}^{-1}$.  The bolometry results on sample~\#2 suggest a much lower $T\!=\!0$ intercept, although the error estimates are substantial.  The differences between sample~\#1 and sample~\#2 are most likely due to the different filling of the CuO chains in the two samples. Sample~\#1 with $T_c=89$~K has an oxygen content of $y=0.993$, while sample~\#2 with $T_c=91~K$ was not so carefully controlled. The extent to which the off-plane oxygen atoms affect the microwave conductivity is unclear and will be the subject of future studies.

Extracting precise values of $\sigma_1^{1D}(\omega\rightarrow 0,T)$ from the \mbox{Ortho-I} bolometry data is difficult since the spectral width of the narrow component is a sizeable fraction of the experiment's bandwidth at all temperatures.  In the \mbox{Ortho-II} case we observe much narrower low $T$ spectra which allow more unambiguous extractions of $\sigma_1^{1D}(\omega\rightarrow 0,T)$. For \mbox{Ortho-II}, $\sigma_1^{1D}(\omega\rightarrow 0,T)$ varies nonlinearly with $T$ and approaches a low $T$ limiting value of $(2\pm1)\times 10^6\,\Omega^{-1} {\rm m}^{-1}$.  Despite these differences, $\sigma_1^{1D}$ does not vanish as $T\rightarrow 0$ for either \mbox{Ortho-I} or \mbox{Ortho-II} \YBCO.   This observation is consistent with the theoretical work of Atkinson\cite{Atkinson} that suggests oxygen vacancies in the CuO chains of \YBCO\ may serve as strong pair breakers.

\begin{table}
\begin{tabular}{|c|c|c|c|}
\hline Doping & Orientation  & RQOS  \\
  &  &  ($10^{17}\,(\Omega {\rm m s})^{-1}$) \\
\hline\hline
Ortho-I & $\hat{a}$ & $6\pm 2$ \\
sample \#1& $\hat{b}$ & $10\pm 7$ \\ \hline
Ortho-I & $\hat{a}$ & $7\pm 1$ \\
sample \#2& $\hat{b}$ & $7.2\pm 0.6$ \\ \hline
Ortho-II & $\hat{a}$ & $6\pm 2$ \\
& $\hat{b}$ & $10\pm 6$ \\ \hline
\end{tabular}
\caption[]
{Residual quasiparticle oscillator strength (RQOS) from the bolometry data shown in Figs.~\ref{fig:figure10} and \ref{fig:figure11}.  At most this corresponds to 3\% of the total spectral weight.}
\label{table:SpectralWeightIntercepts}
\end{table}

\section{Conclusions}

The similarities and differences between the data shown for \OrthoII\
and \Overdoped\ offer insight into the nature of the in-plane
anisotropy in this system, the doping dependence of the inelastic
scattering, and the elastic scattering of nodal quasiparticles by
defects. Both materials show a very broad component in $\sigma(\omega)$ in the
direction parallel to the CuO chains, which is presumably due to the
quasi-one-dimensional Fermi surface sheet derived from a CuO chain
band hybridized with the \cuplane\ plane bands. Atkinson showed that such a
mechanism is needed to explain the anisotropy of the in-plane
superfluid density.\cite{Atkinson} He also needed to invoke strong
pair-breaking by O defects to explain the lack of a low temperature upturn in the $\hat{b}$-axis superfluid density, an upturn expected if the superconductivity on the chains arises from the proximity effect. This strong scattering would likely explain the broad $\sigma(\omega)$ spectrum of quasiparticle excitations on this Fermi surface sheet.

An important and puzzling issue arises by examining the anisotropy of the transport relaxation rate and temperature dependent spectral weight.  In the cuprates, the {\mbox CuO$_2$} plane currents are carried by nodal quasiparticles.  Since the gap nodes lie close to the diagonals of the Brilloiun zone, these excitations carry both $\hat{a}$ and $\hat{b}$-axis currents. We might therefore expect the transport relaxation rate, as measured by the width of the $\sigma_1^{2D}(\omega,T)$ spectra, to be $\hat{a}$--$\hat{b}$ plane isotropic. This is indeed the case in the Ortho-I sample: the conductivity spectra have approximately the same width in the $\hat{a}$ and $\hat{b}$ directions, and the temperature-dependent spectral weights agree to within experimental uncertainty.  The latter conclusion is sensitive to the choice of $\lambda_L(T\to0)$, but we conclude that Ortho-I \Overdoped\ is well described by a 3-band model consisting of a CuO chain-like band acting in parallel with the bonding/antibonding CuO$_2$ plane bands. However, the Ortho-II results present a surprising contrast.  The widths of  the  conductivity spectra and the temperature-dependent quasiparticle spectral weights differ by a factor of two between the $\hat{a}$ and $\hat{b}$ directions.  One of the possible explanations for this anisotropy may be the effect of Ortho-II ordering on the electronic structure of the material.  Ortho-II ordering doubles the unit cell of \YBCO\ in the $\hat{a}$-axis direction, and can therefore substantially alter the Fermi surface topology with respect to the fully-doped compound.  In particular, Bascones {\em et al.} have predicted that Ortho-II ordering can generate two new quasi-one-dimensional bands that are derived entirely from the CuO$_2$ plane state.\cite{Bascones}  This might generate additional nodal quasiparticles derived from the new quasi-one-dimensional bands, increasing the $\hat{a}\!:\!\hat{b}$ anisotropy of the in-plane conductivity. We leave this issue as an open question for future work in this area, and as a feature that must be kept in mind for any theoretical treatment of this phase.

Figure \ref{fig:figure10} demonstrates that the rapid collapse of the
inelastic scattering below $T_c$ occurs in both of these highly-ordered phases of \YBCO. One theoretical proposal treats the opening of a gap in the spectrum of excitations that scatters the nodal quasiparticles, including the key detail that Umklapp processes are needed to relax charge currents.\cite{walker00,duffy01}  The $T$ dependence of the damping $1/\Lambda$ is consistent with exponentially-activated behaviour,
the expectation for Umklapp scattering.  The activated behaviour arises because there is a minimum threshold quasiparticle wave vector that satisfies the Umklapp requirement.
The new results for the Ortho-II phase \OrthoII\ are qualitatively
consistent with this picture. The damping rises even more quickly,
broadening $\sigma_1(\omega)$ out of the microwave range by 20 K.
This could result not only from the lower $T_c$, but also because
changes in the Fermi surface at this lower doping would enhance Umklapp
processes by moving the required excitation closer to the
nodes as compared to \Overdoped. Unfortunately, a quantitative fit is
difficult because of the strong linear term in the elastic scattering in this sample.

Quasiparticle scattering in \Overdoped\ and \OrthoII\ samples is likely dominated by elastic scattering below about 20 and 10~K, respectively.  In both cases, the
lowest temperature spectrum at 1.3 K has the cusp-like shape
expected for weak scattering.\cite{TurnerPRL,schachinger03,Kim}  However, the \OrthoII\ spectra retain the weak-scattering cusp-like form up to higher temperatures where
inelastic scattering takes over, but the \Overdoped\ spectra take on
a more Drude like shape and do not show the temperature independent
$\sigma_1(\omega\rightarrow 0)$ expected for weak limit scattering.
One possible explanation is that the scatterers are neither in the
Born nor the unitary limit, something also suggested by recent
thermal conductivity measurements.\cite{Hill04} In such a case, the
shape and temperature evolution depend on the strength of the
scattering, the density of scatterers and the size of the
superconducting energy gap. Detailed fits to the spectra will be
the subject of future work, but one can qualitatively state that the
scattering in \Overdoped\ appears to be slightly farther from the
Born limit than it is in \OrthoII. It has also been suggested that
the nearly linear temperature dependence of $\sigma_1(\omega, T)$ for any given fixed frequency $\omega>0$ at low T is also a consequence of intermediate scattering phase shifts.\cite{schachinger02,hensen97}

In attempting to generate a microscopic description of conductivity data many researchers have focused on models of quasiparticle scattering from point-like defects.  This is a logical starting point, but it has become apparent that such models are incapable of simultaneously accounting for the observed spectral lineshapes and the significant amount of oscillator strength that remains
uncondensed at 1.2 K, more than would be expected as the
conductivity evolves towards the $T\!=\!0$ limit. Schachinger {\em et al.} tried to invoke intermediate scattering phase shifts for point scatterers to explain experimental results, but did not present a quantitative set of parameters that could
simultaneously describe the spectral shape, its temperature
dependence, and the uncondensed oscillator strength at 1.2 K.\cite{schachinger03} This
problem is widespread and far more severe in other cuprates. In
${\rm Bi_2Sr_2CaCu_2O_{8+\delta}}$ for instance there is much greater
uncondensed oscillator strength observed in both terahertz measurements of
films \cite{corson00} and microwave measurements of single crystals.\cite{lee96}  In these cases, residual oscillator strength as high as 30 \% of the
superfluid density was observed, compared to the few percent seen in
the \YBCO\ samples discussed here. Motivated by the spectroscopic
inhomogeneity observed in scanning tunneling spectroscopy,\cite{pan01} Orenstein \cite{orenstein03} has suggested that this oscillator strength was due to a collective mode in the presence of inhomogeneity. More recently, Nunner {\em et al.} have been
driven by tunneling results to suggest that randomly distributed interstitial O dopants may be the source of the electronic inhomogeneity in ${\rm Bi_2Sr_2CaCu_2O_{8+\delta}}$.\cite{nunner05} This has suggested that rather than focussing on point scatterers,
a better approach is to study extended scatterers, which give rise to slowly varying potentials on the CuO$_2$ planes created by off-plane disorder.
Such models start from knowledge of the actual defects in the
system and have had considerable success in fitting both the
${\rm Bi_2Sr_2CaCu_2O_{8+\delta}}$ microwave data and the \Overdoped\ data
presented here.  The basic suggestion is that off-plane disorder is a
controlling feature of all cuprates and that even though the relative amount
disorder is much smaller in the oxygen-ordered phases of \YBCO, it
still plays an important role in understanding its electronic properties. The
new spectroscopic data presented here, together with that on other
materials\cite{lee96,broun97} provides a testing ground for the
different models: collective modes in an inhomogeneous
superconductor, point scatterers with intermediate phase shifts, and
scattering by off-plane disorder.

\acknowledgments

The authors gratefully acknowledge financial support from the Natural Science and Engineering Research Council of Canada and the Canadian Institute
for Advanced Research.

\end{document}